\documentclass[letterpaper,twocolumn,10pt]{article}

\usepackage[hyphens]{url}
\usepackage{usenix}
\usepackage{cite}
\usepackage{amsmath,amssymb,amsfonts}
\usepackage{algorithm}
\usepackage{algpseudocode}
\usepackage{graphicx}
\usepackage{textcomp}
\usepackage{xcolor}
\usepackage{enumitem}
\usepackage{multirow}
\usepackage{float}
\usepackage[font=footnotesize]{caption}
\usepackage{subcaption}

\newcommand{\changes}{\textcolor{black}}

\usepackage{etoolbox}
\makeatletter
\patchcmd{\maketitle}
	{\@maketitle}
	{\@maketitle\vspace{-8em}}% change the value as needed
	{}
	{}
\makeatother

\def\BibTeX{{\rm B\kern-.05em{\sc i\kern-.025em b}\kern-.08em
    T\kern-.1667em\lower.7ex\hbox{E}\kern-.125emX}}
\begin{document}

\newcommand{\qed}{\hfill$\blacksquare$}

% \title{Multi-Factor Key Derivation Function (MFKDF)}
\title{\vspace{-6em}Multi-Factor Key Derivation Function (MFKDF)\\for Fast, Flexible, Secure, \& Practical Key Management\vspace{-1em}}

\author{
{\rm Vivek Nair and Dawn Song}\\
University of California, Berkeley\\
% copy the following lines to add more authors
% \and
% {\rm Name}\\
%Name Institution
} % end author

% \author{\IEEEauthorblockN{Vivek Nair}
% \IEEEauthorblockA{
% \textit{UC Berkeley}\\
% vcn@berkeley.edu}
% \and
% \IEEEauthorblockN{Dawn Song}
% \IEEEauthorblockA{
% \textit{UC Berkeley}\\
% dawnsong@berkeley.edu}
% }

% \author{\IEEEauthorblockN{}
% \IEEEauthorblockA{
% \textit{}\\
% }
% \and
% \IEEEauthorblockN{}
% \IEEEauthorblockA{
% \textit{}\\
% }
% }

\maketitle

\begin{abstract}
We present the first general construction of a Multi-Factor Key Derivation Function (MFKDF). Our function expands upon password-based key derivation functions (PBKDFs) with support for using other popular authentication factors like TOTP, HOTP, and hardware tokens in the key derivation process. In doing so, it provides an exponential security improvement over PBKDFs with less than 12~ms of additional computational overhead in a typical web browser. We further present a threshold MFKDF construction, allowing for client-side key recovery and reconstitution if a factor is lost. Finally, by ``stacking'' derived keys, we provide a means of cryptographically enforcing arbitrarily specific key derivation policies.
The result is a paradigm shift toward direct cryptographic protection of user data using all available authentication factors, with no noticeable change to the user experience.
We demonstrate the ability of our solution to not only significantly improve the security of existing systems implementing PBKDFs, but also to enable new applications where PBKDFs would not be considered a feasible approach.
\end{abstract}

% \begin{IEEEkeywords}
% Key derivation, Account recovery, Multi-factor authentication, HOTP, TOTP, Applied cryptography
% \end{IEEEkeywords}

\section{Introduction}
\label{sec:introduction}
Since the introduction of PBKDF1 in 1991, password-based key derivation functions (PBKDFs) have enjoyed widespread use and deployment in a variety of settings, including in the WPA~\cite{wpa} and WPA2~\cite{wpa2} protocols, LastPass~\cite{lastpass_arch} and Dashlane~\cite{dashlane_arch} applications, and Windows~\cite{windows} and iOS~\cite{ios} operating systems. PBKDFs provide a convenient solution to the usable key management problem: by deterministically deriving keys based on a user's password, systems can achieve end-to-end encryption while providing a seamless user experience and subverting the need for secure key storage. However, the recent surge in password-based attacks like credential stuffing and password spraying has highlighted the critical weakness of passwords as a sole authentication factor.

Although a growing number of platforms have implemented multi-factor authentication in response to this threat, password-derived keys (and thus all secrets encrypted with them) remain only as secure as the passwords they are based on. We therefore suggest the use of, and provide a novel construction for, a key derivation function that incorporates all of a user's multiple authentication factors into the key derivation process. In doing so, user data is, for the first time, directly cryptographically protected by all authentication factors.

While incorporating all of a user's login factors into the key derivation process seems like a natural and long overdue extension of PBKDFs, achieving this with the most popular secondary authentication factors currently in use is difficult in practice. Current PBKDFs rely on the relatively unchanging nature of passwords to convert them, via deterministic one-way functions (OWFs), into fixed encryption keys. By contrast, most of the secondary factors in popular use today constitute one-time passwords (OTPs) that are expected to change upon each user login, which does not readily facilitate the derivation of a static key.

In this paper, we describe and evaluate the first known Multi-Factor Key Derivation Function (MFKDF) with support for common authentication factors like TOTP, HOTP, OOBA (e.g., Email/SMS), HMAC-SHA1 (e.g., YubiKey), and, of course, static factors like passwords and recovery codes. We do so using fast, standardized cryptographic primitives, resulting in a total computational overhead of $\leq 12$~ms over PBKDFs in a typical web application (\S\ref{sec:performance}), with no noticeable changes to the user experience.

Our MFKDF construction represents a fast, flexible, secure, and practical key management solution that provides several advantages over existing PBKDFs: it supports self-service client-side key recovery without creating a central point of failure, can be used to implicitly authenticate without revealing authentication factors to a server, and enables cryptographic enforcement of arbitrarily complex authentication policies. We demonstrate these features by implementing and testing MFKDF in realistic, full-stack centralized and decentralized applications, including a trustless decentralized cryptocurrency wallet for which PBKDFs would not have been sufficiently secure (\S\ref{sec:decentralized}).\\

\noindent \textbf{Contributions}
\begin{enumerate}[leftmargin=*]
    \itemsep -0.2em
    \item We provide the first general construction of a multi-factor key derivation function (\S\ref{sec:mfkdf}). Our function provides an exponential security improvement over PBKDFs (\S\ref{sec:entropy}).
    \item We provide KDF constructions for several popular authentication factors (\S\ref{sec:factors}), including the first known KDF constructions based on HOTP and TOTP (\S\ref{sec:soft}).
    \item We provide a $k$-of-$n$ threshold MFKDF construction (\S\ref{sec:tmfkdf}) that can be used to facilitate self-service account recovery without a central point of failure (\S\ref{sec:recovery}).
    \item We illustrate how MFKDF can cryptographically enforce arbitrarily specific key derivation policies (\S\ref{sec:policy}).
\end{enumerate}

\clearpage
\section{Background \& Motivation}
\label{sec:background}
\vspace{-0.6em}
In this section, we aim to motivate the need for a multi-factor key derivation function by following the prototypical use case of a password management system. Password managers represent a common application for client-side key derivation, with password-based key derivation functions currently being a major architectural component of popular password management products like LastPass \cite{lastpass_arch}, Dashlane \cite{dashlane_arch}, and 1Password \cite{1password_arch}. We first describe a typical architecture of a modern password management system, then describe the drawbacks of this current architecture, and finally illustrate how the multi-factor key derivation function presented herein can serve to remedy these flaws. \\
\vspace{-0.5em}

\noindent \textbf{Password-Based Key Derivation}\vspace{0.2em}\\
Password management systems are, by their nature, highly security-sensitive due to their role as a gateway to all of a user's online accounts. Systems that provide adequate security for this application are generally trustless, as users are often concerned not only with the threat of outside actors, but also with potential malicious activity of the service itself. The chief goal of a secure password management system architecture is therefore to preserve the confidentiality of stored secrets even in the event that all centralized system components are compromised by an adversary.\footnote{To further motivate this standard, we note that LastPass has experienced at least 6 security incidents \cite{lastpass_security}, including two database breaches \cite{siegrist_lastpass_2015, toubba_notice_2022}.} Today, password-based key derivation functions like PBKDF1 and PBKDF2 \cite{rfc2898} are critical tools for enabling this level of security. In general, a password-based key derivation function $F$ converts a password $P$, salt $S$, and optional configuration parameters $\mathit{cfg}$ into a key $DK$ of length $\mathit{dkLen}$: $DK=F(P,S,\mathit{cfg},\mathit{dkLen})$.

In practice, most password-based key derivation functions are built upon, and inherit the security properties of, cryptographic hash functions like SHA-256 \cite{rfc6234}. When a user signs in to a typical password management application, PBKDF2 is used on the client side to derive a key from the user's password. A symmetric encryption function like AES-256 \cite{fips197} is then used to encrypt all of a user's secrets on the client side prior to their storage in a centralized database. Thus, even an adversary with complete access to the database will not be able to derive the key necessary to decrypt the user's secrets without knowing their password.

An important property of password-based key derivation functions is a degree of intentional computational inefficiency that increases the relative difficulty of brute-force attacks. For example, the PBKDF2 configuration used by LastPass invokes 100,000 rounds of SHA-256 to increase its computational difficulty to a degree that remains relatively unnoticeable to users but is significantly burdensome to brute-force attackers. Advanced password-based key derivation functions like Argon2 \cite{argon2} have been developed to further resist brute-force attacks, but operate in a fundamentally similar way to the functions like PBKDF2 described above.

\noindent \textbf{Multi-Factor Authentication}\vspace{0.2em}\\
While the password-based key derivation approach described above is effective at binding a user's secrets to their master password, it is not adequate on its own to protect a user's account due to the well-known insecurity of passwords as a sole authentication factor \cite{password_reuse, florencio_large_2006} and their susceptibility to attacks such as credential stuffing \cite{credential_stuffing}. Therefore, services typically use multi-factor authentication (MFA) in conjunction with password-based key derivation. Popular secondary authentication factors include ``soft tokens'' like HMAC-based One-Time Password (HOTP) \cite{rfc4226} and Time-based One-Time Password (TOTP) \cite{rfc6238}, ``hard tokens'' like YubiKeys \cite{yubikey}, and Out-of-Band Authentication (OOBA) factors like email and SMS \cite{ooba}. These factors are inserted by password managers into the login process for obtaining an authentication token necessary to access encrypted secrets stored in a database.

The use of MFA during the login process may prevent attackers from accessing stored secrets under correct system operation, but fails to meet the previously-stated security goal of surviving a complete system breach. In the event of a compromise, passwords remain the only factor necessary to decrypt and access secrets. Thus, the current method of MFA only superficially addresses the threat of attacks like password spraying and credential stuffing in the context of password management. What is therefore needed is a mechanism for ensuring that a user's encryption key cannot be derived without utilizing all of their authentication factors.

\medskip

\begin{figure}[h]
\includegraphics[width=0.65 \linewidth]{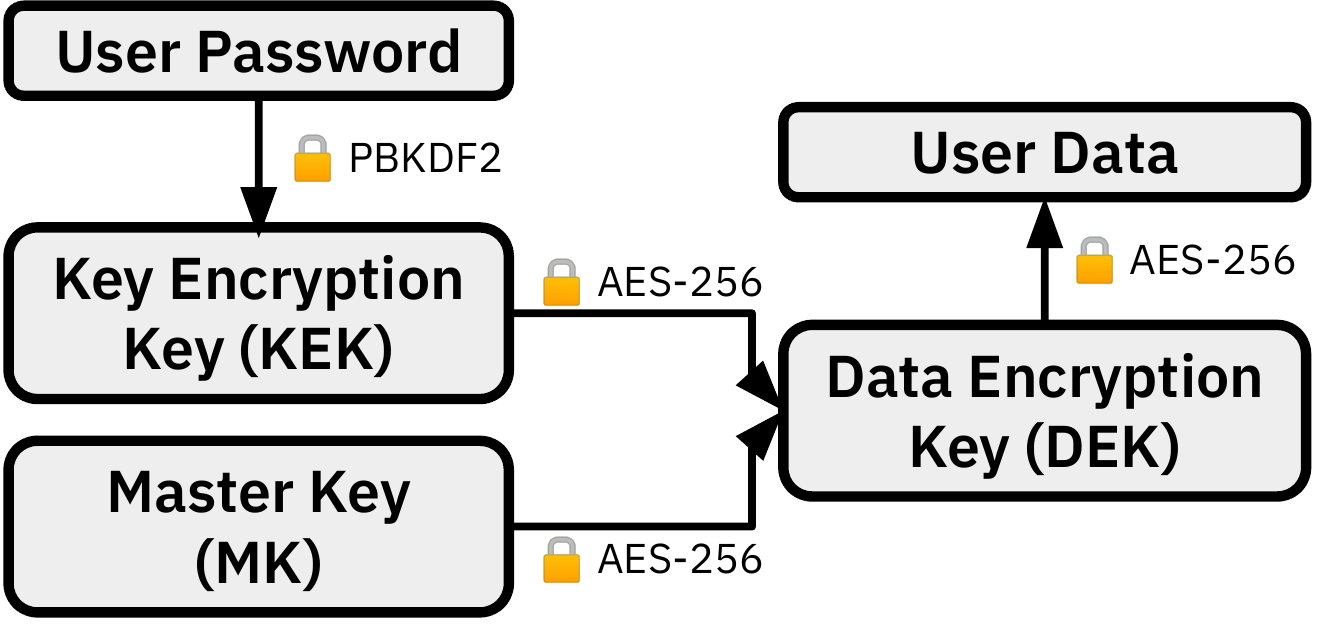}
\centering
\vspace{-0.5em}
\caption{Typical architecture for account recovery via master key.}
\label{fig:nist_sp80057}
\end{figure}

\noindent \textbf{Account Recovery}\vspace{0.2em}\\
Absent additional considerations for account recovery, systems using password-based key derivation are liable to a complete loss of user data in the event of a lost password. In light of the fact that user passwords are, in fact, frequently forgotten by end users \cite{password_reset}, this risk is generally considered untenable for users and service providers alike. This risk can be mitigated via the use of a master key as shown in Fig. \ref{fig:nist_sp80057}.

\vspace{-0.25em}
\vspace{-0.75em}

As illustrated above, when a user forgets their password, they are presented with one or more alternative security challenges (e.g., security questions). Upon successful completion of these challenges, the system uses the master key to recover the data encryption key and decrypt any stored data.

While the use of a central master key for data recovery may be deemed a necessary concession for the sake of usability, it once again fails to meet our security goal of surviving centralized system compromise; in fact, it largely defeats the initial purpose of using client-side user-derived keys.

Due to the obvious drawbacks of this approach, modern password managers rarely rely on central recovery keys for individual accounts.
Instead, they largely utilize approaches that store recovery keys on trusted user devices. For instance, LastPass \cite{lastpass_arch}, 1Password \cite{1password_arch}, and Dashlane \cite{dashlane_arch} all support account recovery via biometrics by storing a key on a trusted smartphone. LastPass additionally supports SMS recovery by storing a key on a trusted browser, while 1Password supports the use of Linux authentication for recovery.

In all instances, without a central key, users must maintain access to at least one trusted device where a recovery key has previously been stored if they hope to recover their account. Some services, like 1Password, go as far as to allow users to manually store and manage their secret key, which must be used together with their password to authenticate on new devices. Once again, no recourse is available if this key is not retained in at least one secure location.
What is needed is a key derivation method that supports client-side recovery of secrets when one or more authentication factors is forgotten while requiring neither a trusted device nor a central point of failure. \\

\vspace{-0.8em}

\noindent \textbf{Motivation}\vspace{0em}\\
In summary, current system architectures based on password-based key derivation are flawed in a number of significant ways. They fail to incorporate secondary authentication factors into the key derivation process, leaving users susceptible to credential stuffing and password spraying, and introduce a central point of failure when using master keys for account recovery. These flaws are not problems of specific system architectural designs, but rather are fundamentally implicated by the use of password-based key derivation in a system.

In this paper, we aim to present a complete solution to the problems highlighted above by constructing a multi-factor key derivation function (MFKDF) that realizes the full benefits of multi-factor authentication within the key-derivation process (\S\ref{sec:mfkdf}). Unlike current architectures, our suggested approach supports client-side key recovery without using a master key or trusted device (\S\ref{sec:recovery}). In \S\ref{sec:centralized}, we demonstrate a proof-of-concept password management system architecture based on MFKDF and illustrate its security advantages over current architectures. Importantly, we do so with no noticeable change to the user experience, and while supporting all of the same 2FA factors that users have already grown accustomed to, providing a more streamlined account login and recovery experience than systems using manual key management.

Of course, the current applications of password-based key derivation go far beyond password management, including use in the Windows \cite{windows} and iOS \cite{ios} operating systems and WPA \cite{wpa} and WPA2 \cite{wpa2} wireless protocols. The MFKDF approach presented herein is naturally applicable to many of these systems, such as to bring the benefits of multi-factor authentication to operating systems and wireless networks. Further, in \S\ref{sec:decentralized}, we will illustrate how the security properties of MFKDF enable its use even in contexts where PBKDFs would never have been considered sufficiently secure.
\section{Problem Statement}
\label{sec:setup}

Krawczyk's analysis of HKDF \cite{hutchison_cryptographic_2010} provides an excellent formalization of key derivation functions. We use a modified framework of Krawczyk here to give a formal statement of the security of our MFKDF constructions. \\

There are two components to our framework: \textit{factor constructions}, which exist to derive static key material from a specific authentication factor, and \textit{KDF constructions}, which use the results of one or more factor constructions to derive a key. We begin by discussing the factor constructions. \\

\noindent \textbf{Factor Constructions}\vspace{0.2em}\\
\noindent \textbf{Definition 1.} A factor $F$ is a two-valued probability distribution $(\sigma, \alpha)$ generated by an efficient probabilistic algorithm.

\medskip

Factors are said to have a private component ($\sigma$) and a public component ($\alpha$). For example, if $F=(\sigma,\alpha)$ is a security question, then $\alpha$ might contain the publicly-known question and $\sigma$ might contain the secret answer. On the other hand, if $F$ is a password, then $\alpha$ might contain the publicly-known strength requirements while $\sigma$ contains the password itself.
In this paper, we refer to the public component $\alpha$ as the ``factor parameters'' and the private component $\sigma$ as the ``factor material.'' For some factors, the parameters $\alpha$ might change over time, but the material $\sigma$ must remain constant.

\medskip

\noindent \textbf{Definition 2.} We say that $F$ is a statistical $m$-entropy factor if for all $s$ and $a$ in support of $F$, the conditional probability $\mathit{Prob}(\sigma=s\mid\alpha=a)$ is at most $2^{-m}$.

\medskip

Consider a factor $F$ where $\alpha$ is always \textit{``What's your favorite number between 1 and 64?''} and users chose $\sigma$ uniformly in $[1,64]$. Then $F$ is a statistical 6-entropy factor.

\medskip

\noindent \textbf{Definition 3.} We say that $F$ is a computational $m$-entropy factor if there is a statistical $m$-entropy factor $F'$ that is computationally indistinguishable from F.

\medskip

Consider modifying the previous example such that $\alpha$ also contains a secure encryption of $\sigma$. Now $F$ is a computational 6-entropy factor since $\sigma$ is theoretically determinable within $[1,64]$ given $\alpha$, yet it is computationally infeasible to do so (compared to the difficulty of guessing $\sigma \in [1,64]$).

\medskip

\noindent \textbf{Definition 4.} A factor construction $\mathcal{F}$ uses a factor witness $W_i$ and parameters $\alpha_i$ to output the factor material $\sigma$ and new parameters $\alpha_{i+1}$; ${\mathcal{F}: (W_i,\alpha_i) \mapsto (\sigma, \alpha_{i+1})}$.

\medskip

A factor witness ($W_i$) refers to the one-time value used to demonstrate possession of a factor. For example, $W_i$ might be a one-time 6-digit code in the case of HOTP. The role of the factor construction $\mathcal{F}$ is to output a constant $\sigma$ despite the witness $W_i$ potentially changing. The ability to update the parameters ($\alpha_i \mapsto \alpha_{i+1}$) each time $\mathcal{F}$ is invoked is critical to making this possible.

In this paper, we split $\mathcal{F}$ into the $\mathit{FactorDerive}$ function, producing $\sigma$, and the $\mathit{FactorUpdate}$, producing $\alpha_{i+1}$. Each factor also requires a corresponding $\mathit{FactorSetup}$ function to produce the initial parameters $\alpha_{0}$ given some configuration.

\medskip

\noindent \textbf{Goals.} In this paper, we aim to provide factor constructions ($\mathit{FactorSetup}$, $\mathit{FactorDerive}$, and $\mathit{FactorUpdate}$) for computational $m$-entropy factors representing HOTP, TOTP, \mbox{YubiKey}, OOBA, and constant factors like passwords. \\

\noindent \textbf{KDF Constructions}\vspace{0.2em}\\
\noindent \textbf{Definition 5.} A KDF $\mathcal{D}$ outputs $K$ (a string of $\ell$ bits) and $\alpha_{i+1}$ given $\ell$ and values $(\sigma, \alpha_i)$ sampled from a factor $F$.

\medskip

Note that by definition, a KDF is defined with respect to a single factor. In practice, we will show that $F$ can be a ``composite factor'' which combines multiple input factors, allowing KDF to be a multi-factor KDF (MFKDF). Thus, our MFKDF construction needs both an $\mathit{MFKDFDerive}$ function ($\mathcal{D}$) and an $\mathit{MFKDFSetup}$ function that establishes the initial parameters ($\alpha_0$) of the composite factor $F$.

\medskip

\noindent \textbf{Definition 6.} A $\mathit{KDF}$ is said to be $(t,q,\varepsilon)$-secure with respect to a factor $F$ if given a $(\sigma, \alpha) \in F$ no attacker $\mathcal{A}$ knowing $\alpha$ running in time $t$ and making at most $q$ arbitrary queries to $\mathit{KDF}$ can distinguish between $\mathit{KDF}(\sigma,\ell)$ and $\{0,1\}^\ell$ with probability larger than $1/2+\varepsilon$.

\medskip

Def. 6 is our main security definition for MFKDF. It states that the output key $K$ should be hard to distinguish from random noise $\{0,1\}^\ell$ without knowing the secret component (factor material) $\sigma$ of the input factor $F$.

\medskip

\noindent \textbf{Definition 7.} A KDF is called $(t,q,\varepsilon)$ $m$-entropy secure if it is $(t,q,\varepsilon)$-secure with respect to all $m$-entropy factors.

\medskip

Note that because our MFKDF constructions require special treatment of each factor type, we cannot assert their security on arbitrary $m$-entropy factors. However, it is reasonable to make this assumption of the underlying KDF (e.g., HKDF~\cite{hutchison_cryptographic_2010}) used to construct MFKDF.

\bigskip

\noindent \textbf{Goals.} In this paper, we aim to provide an MFKDF construction ($\mathit{MFKDFSetup}$, $\mathit{MFKDFDerive}$) which uses a set of factors ${S=\{F_A,F_B,...\}}$ and parameters $\alpha_i$ to produce a key $K$ and new parameters $\alpha_{i+1}$; $\mathcal{D}: (S,\alpha_i) \mapsto (K, \alpha_{i+1})$.

We also aim to provide a $k$-of-$n$ threshold multi-factor KDF construction, where given any $C$ consisting of $\geq k$ factors in $S$ (\mbox{$\forall C \subseteq S$ s.t. $|C| \geq k$}), $\mathcal{D}: (C,\alpha_i) \mapsto (K, \alpha_{i+1})$. \\

\noindent \textbf{Problem Setting}\vspace{0.2em}\\
Per the above definitions, each $\mathit{KDF}$ and factor $F$ has private components ($W_i$, $\sigma$, $K$) and public components ($\alpha$). We assume a legitimate user is the only party able to generate factor witnesses $W_i$ for their authentication factors, and thus is the only party able to derive factor material $\sigma$ and keys $K$. We make no assumptions, however, of parties knowing $\alpha$; the parameters $\alpha$ can even be stored on a public blockchain.

\medskip

\noindent \textbf{Definition 8.} Factors $F_A$ and $F_B$ are independent if for all $(\sigma_a, \alpha_a) \in F_A$ and $(\sigma_b, \alpha_b) \in F_B$, ${P(\sigma_a=s_a \mid \alpha_a=a_a)=P(\sigma_a=s_a \mid \alpha_a=a_a \land \alpha_b=a_b)}$, $\land~ {P(\sigma_b=s_b \mid \alpha_b=a_b)=P(\sigma_b=s_b \mid \alpha_a=a_a \land \alpha_b=a_b)}$.

We add the assumption of independent factors to our problem setting and security definitions. Most factors in practice are independent because they rely on a random key. A counter-example would be a security question whose answer overlaps with the contents of a password. \\

\noindent \textbf{Policy Constructions}\vspace{0.2em}\\
Finally, we introduce the notion of a ``policy-based MFKDF,'' a stronger construction than the ones introduced above which allows implementers to specify exactly which factor combinations are allowed to derive a key.

\medskip

\noindent \textbf{Definition 9.} Given a set of factors ${S=\{F_A,F_B,...\}}$, an allowable factor combination $C$ is any subset of factors in $S$ (${C \subseteq S, C \neq \varnothing}$) that can be used to derive a key $K$.

\medskip

\noindent \textbf{Definition 10.} A policy $P$ is a set of all allowable factor combinations ($P=\{C_1,C_2,...\}$) that can be used to derive a key $K$. \newline

In the previously-described $k$-of-$n$ threshold MFKDF, $P$ was restricted to containing all subsets of $S$ of size $\geq k$. However, per the above definition, $P$ can now contain any non-empty subsets of $S$, regardless of size.

\medskip

\noindent \textbf{Goals.} We aim to provide a policy-based multi-factor KDF construction w.r.t. $P$: \mbox{$\forall C \in P$}, $\mathcal{D}_P: (C,\alpha_i) \mapsto (K, \alpha_{i+1})$.

\medskip

\noindent \textbf{Security Goals.} Standard and threshold MFKDF are sub-cases of policy-based MFKDF, so we define our security goals in the policy setting. Let ${S=\{F_1,F_2,...,F_n\}}$ be $n$ independent computational ${\{m_1,m_2,...,m_n\}}$-entropy factors. Let $\mathit{PKDF}$ be $(t,q,\varepsilon)$-secure. Now given a policy-based MFKDF $\mathit{KDF}$ w.r.t. $\mathit{PKDF}$ and any $P \in \mathcal{P}(\mathcal{P}(S) \setminus \varnothing) \setminus \varnothing$:
% Let ${F_K=(\forall \sigma \in S, \forall \alpha \in S)}$ be a computational $\Sigma S$-entropy source. 
% Then $n$-of-$n$ MFKDF is $(t,q,\varepsilon)$-secure with respect to $F_K$.

\begin{itemize}[leftmargin=*]
    \itemsep -0.2em
    \item \textbf{Correctness.} For any \mbox{$C \in P$}, $(K,\alpha') \gets \mathit{KDF}_P(C,\alpha)$, and \mbox{$C' \in P$}, $(K',\alpha'') \gets \mathit{KDF}_P(C',\alpha')$, $K=K'$. \\
    \emph{(Providing any valid combinations of factors should always result in deriving the same key.)} \medskip
    
    \item \textbf{Safety.} For any \mbox{$C \in P$}, $(K,\alpha') \gets \mathit{KDF}_P(C,\alpha)$, and \mbox{$C' \not\in P$}, $(K',\alpha'') \gets \mathit{KDF}_P(C',\alpha')$, $K \neq K'$ except with negligible probability w.r.t. key size $\ell$. \\
    \emph{(Providing an invalid set of factors should be highly unlikely to derive the correct key.)} \medskip
    
    \item \textbf{Entropy}. Let $E$ denote $\sum\{m_1,m_2,...,m_n\}$ for all ${C \in P}$. Then the MFKDF construction w.r.t. $P$ shall be \mbox{$(t,q,\varepsilon)$-secure} w.r.t. a factor $F_K$, where $F_K$ is a computational \mbox{($min(E)$)-entropy} factor. \\
    \emph{(Attacking the derived key should be as hard as attacking the weakest set of allowed factors.)}
\end{itemize}

% With these goals in mind, we begin by presenting a basic $n$-of-$n$ MFKDF in \S\ref{sec:mfkdf}, then provide $m$-entropy factor constructions in \S\ref{sec:factors}. The $k$-of-$n$ threshold MFKDF is described in \S\ref{sec:threshold}, and finally, the policy-based MFKDF is given in \S\ref{sec:policy}.
\section{Multi-Factor Key Derivation}
\label{sec:overview}

We identified a few fundamental difficulties with implementing even a basic $n$-of-$n$ multi-factor key derivation function. Firstly, there are a wide variety of authentication factors that must be supported, each requiring its own treatment of inputs and outputs to be used for key derivation. Secondly, many of these factors are constantly-changing OTPs, which nevertheless need to be used to derive the correct, static key if and only if the correct OTP (i.e., witness $W_i$) is provided at that instant. A third, self-imposed constraint is ``plug-and-play'' compatibility with existing systems using PBKDFs; systems should not need to be entirely rearchitected to use MFKDF, and the user experience should not be impacted whatsoever. This section outlines, at a high level, an MFKDF design that meets these goals along with the security goals of \S\ref{sec:setup}. A formal statement of this design is given in \S\ref{sec:mfkdf_construction}.

\subsubsection*{MFKDF Construction}
\label{sec:mfkdf}

\begin{figure}[H]
\includegraphics[width=0.97 \linewidth]{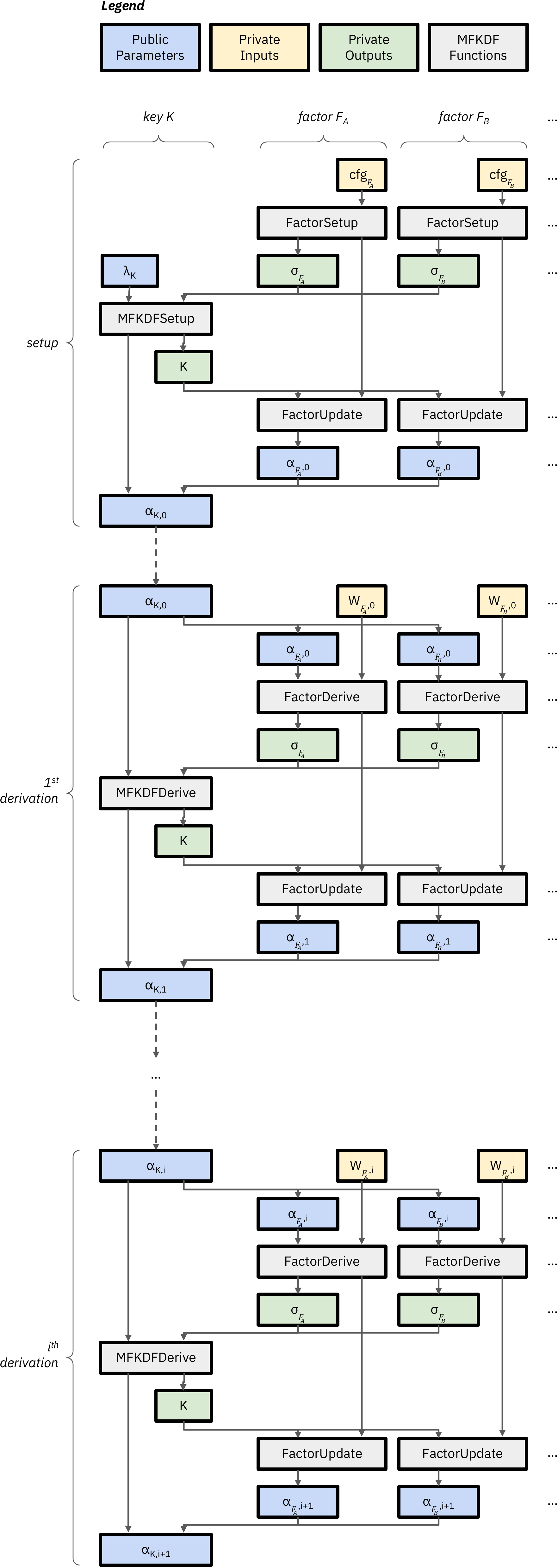}
\centering
\caption{Multi-phase modular MFKDF construction.}
\label{fig:main}
\end{figure}

In Fig. \ref{fig:main}, we illustrate an MFKDF system consisting of general-purpose $\mathit{MFKDFSetup}$ and $\mathit{MFKDFDerive}$ functions, which present a standard interface to any number of factor-specific $\mathit{FactorSetup}$, $\mathit{FactorDerive}$, and $\mathit{FactorUpdate}$ functions. This modular approach allows for individual treatment of various authentication factors, many of which are described in \S\ref{sec:factors}, as well as potential forward-compatibility with future factors. During a setup phase (e.g., upon account creation), initial key parameters ($\alpha_{K,0}$) are produced, encapsulating several factor parameters $\{\alpha_{\mathit{F_A},0}, \alpha_{\mathit{F_B},0}, ...\}$. The $i$th derive phase (upon $i$th login) then proceeds as follows:

\begin{enumerate}[leftmargin=*]
    \itemsep -0.2em
    \item The key parameters $\alpha_{K,i}$ are split into several factor parameters $\{\alpha_{F_A,i}, \alpha_{F_B,i}, ...\}$.
    \item Each factor witness $\{W_{F_A,i}, W_{F_B,i}, ...\}$, along with its factor parameters, $\{\alpha_{F_A,i}, \alpha_{F_B,i}, ...\}$, is converted by its $\mathit{FactorDerive}$ function into factor material $\sigma_F$.
    \item The factor material $\{\sigma_{F_A}, \sigma_{F_B}, ...\}$ is combined by $\mathit{MFKDFDerive}$ into key material $\sigma_K$.
    \item Let $\mathit{KDF}$ be a memory-hard PBKDF like Argon2 \cite{argon2}. $\mathit{MFKDFDerive}$ uses $\mathit{KDF}$ on $\sigma_K$ to produce key $K$.
    \item The $\mathit{FactorUpdate}$ functions use $K$ and $\alpha_{F,i}$ to produce $\alpha_{F,i+1}$ for each factor.
    \item The factor parameters $\{\alpha_{F_A,i+1}, \alpha_{F_B,i+1}, ...\}$ are combined with the output of $\mathit{MFKDFDerive}$ to produce the updated key parameters, $\alpha_{K,i+1}$.
\end{enumerate}

A major innovation in this approach is the feedback loop that allows the factor constructions to use $K$ in producing $\alpha_{\mathit{F},i+1}$, which has the effect of allowing factor constructions to hide secrets using $K$. These secrets can be used to ``setup'' the ${i+1}$th key derivation during the $i$th derivation, making possible the use of OTP factors.

% This approach achieves plug-and-play compatibility with existing PBKDF-based systems, with $\mathit{PBKDF}$ replaced by the $\mathit{MFKDFDerive}$ and $\mathit{FactorDerive}$ functions, password inputs being replaced by multiple factor witnesses $\{W_{F_A}, W_{F_B}, ...\}$, and $\alpha_K$ replacing the role of a $\mathit{salt}$.

\section{Factor Constructions}
\label{sec:factors}

As described in \S\ref{sec:introduction}, the chief contribution of MFKDF is not necessarily the concept of a KDF with multiple inputs, but rather its support for existing, commonly used authentication factors, including dynamic factors like HOTP/TOTP. In this section, we describe MFKDF factor constructions for a wide variety of popular authentication factors.
The goal of each factor derivation function is to convert factor parameters ($\alpha_{F,i}$) and a factor witness $W_{F,i}$ (e.g., a password, YubiKey response, or TOTP code) into fixed factor material $\sigma_F$ that can be used by MFKDF to derive a key ($K$), along with updated factor parameters ($\alpha_{F,i+1}$). Due to the key-feedback mechanism of \S\ref{sec:mfkdf}, the factor \mbox{derivation} function can use $K$ when producing $\alpha_{F,i+1}$, but not when producing $\sigma_F$. A formal statement of each factor is given in \S\ref{sec:factor_constructions}.

\subsection{Constant Factors}
\label{sec:constant}
We begin by discussing constant factors such as passwords, security questions, and recovery codes, as these are amongst the easiest factors to construct. In general, PBKDFs are already well suited to handle such factors, so the witness $W_{F,i}$ can be directly returned as $\sigma_F$. In some cases, a $\mathit{transform}$ function first is applied to $W_{F,i}$, such as to standardize the case of a security answer, or to extract useful data from a UUIDv4 recovery code. No parameters are required ($\alpha_{F}=\varepsilon$) due to the artifacts being static. Our MFKDF construction would therefore probably be overkill for simply combining multiple static factors, but their inclusion here is useful to facilitate classic factor combinations like password + TOTP.

\subsection{Software Tokens}
\label{sec:soft}

We next describe factor constructions for ``soft tokens,'' specifically HOTP \cite{rfc4226} and TOTP \cite{rfc6238}. Our constructions rely on the information-theoretic security of modular arithmetic rings to convert a dynamic OTP ($W_{F,i}$) into a fixed integer ($\mathit{target}_F$) that can be used as $\sigma_F$. The resulting factors are perfectly backward-compatible with existing HOTP/TOTP implementations like ``Google Authenticator.'' Both support a variable number of digits $d$; typically, $d=6$.

\subsubsection{HOTP}
The HOTP factor construction is the first to take advantage of the key feedback loop described in \S\ref{sec:mfkdf}. Although a user's HOTP code is expected to be different upon each derivation, one can predict what the user's next HOTP code will be if they know the HOTP key ($\mathit{hotpkey}_F$) and counter ($\mathit{ctr}_F$). On the other hand, $\mathit{hotpkey}_F$ cannot be stored openly, as it would allow an attacker to bypass the HOTP factor. The key feedback mechanism allows $\mathit{hotpkey}_F$ to be stored securely in $\alpha_F$ encrypted with the final derived key $K$ while still being used during the $i$th HOTP factor derivation to set up the ${i+1}$th derivation.
During the HOTP factor setup process for $\mathit{hotpkey}_F$, a fixed $\mathit{target}_F$ is set to a random integer in range $[0,10^d)$. The first OTP ($\mathit{otp}_{F,0}$) is determined using $\mathit{hotpkey}_F$ with $\mathit{ctr_{F,0}}=1$, and the modular difference is stored as ${\mathit{offset}_{F,0}=(\mathit{target}_F-\mathit{otp}_{F,0})~\%~10^d}$. The factor parameters $\alpha_{F,0}$ consist of $(d,\mathit{ctr_{F,0}},\mathit{offset}_{F,0},\mathit{ct}_F)$, where $\mathit{ct}_F$ is $\mathit{hotpkey}_F$ encrypted under $K$. The factor derivation process (with $W_{F,i}=\mathit{otp}_{F,i}$) is as follows:

\medskip

\begin{enumerate}[leftmargin=*]
    \itemsep -0.2em
    \item $(d,\mathit{ctr_{F,i}},\mathit{offset}_{F,i},\mathit{ct}_i)$ are obtained from $\alpha_{F,i}$.
    \item ${\sigma_F=\mathit{target}_F}$ is found as ${(\mathit{offset}_{F,i}+W_{F,i})~\%~10^d}$.
\end{enumerate}

\medskip

At this point, the factor material ($\sigma_F$) has been successfully recovered (assuming $W_{F,i}$ is correct), and the MFKDF derivation of $K$ can proceed. The HOTP factor can then use $K$ for the remainder of the derivation process via the key-feedback loop described in \S\ref{sec:mfkdf}:

\medskip

\begin{enumerate}[leftmargin=*]
    \itemsep -0.2em
  \setcounter{enumi}{2}
    \item Decrypt $\mathit{hotpkey}_F$ from $\mathit{ct}_F$ using $K$.
    \item Increment $\mathit{ctr}_F$ ($\mathit{ctr}_{F,i+1}=\mathit{ctr}_{F,i}+1$).
    \item Determine $\mathit{otp}_{F,i+1}$ using $\mathit{hotpkey}_F$ and $\mathit{ctr}_{F,i+1}$.
    \item Calculate $\mathit{offset}_{F,i+1}$ as ${(\mathit{target}_F-\mathit{otp}_{F,i+1})~\%~10^d}$.
    \item Return $(d,\mathit{ctr_{F,i+1}},\mathit{offset}_{F,i+1},\mathit{ct}_F)$ as $\alpha_{F,i+1}$.
\end{enumerate}

\medskip

In summary, the above HOTP factor construction succeeds at converting a dynamic HOTP code into fixed factor material $\sigma_F$ through use of a modular $\mathit{offset}_F$ that is updated each round using $\mathit{hotpkey}_F$ (stored as $\mathit{ct}_F$).

\subsubsection{TOTP}
Our TOTP factor construction uses a similar fundamental approach to the HOTP construction, given that by definition, ${\mathit{TOTP}(K)=\mathit{HOTP}(K,\lfloor(T-T_0)/T_X\rfloor)}$ where $T$ is the current UNIX time, $T_0$ is the initial time, and $T_X$ is the time interval (usually 30 seconds). However, given that we cannot predict exactly which time $T$ the next derivation will occur, each possible $\mathit{offset}_F$ must be calculated within a fixed window $w$, corresponding to $\{\mathit{ctr},\mathit{ctr}+1,...,\mathit{ctr}+t\}$. \changes{Because $\mathit{hotpkey}_F$ is known to the client during the setup and update phases, these offsets can be calculated without involving the TOTP application.} Thus, upon derivation at time $T$, all subsequent derivations between $T$ and $w \cdot T_X$ will succeed at recovering $\sigma_F$ (the same approach can also be used to facilitate counter desynchronization when using HOTP). This approach is quite practical, with a window of ${w=87600}$ (30 days) requiring just $219$~kb of storage, and less than $850$~ms to setup and derive \mbox{(see \S\ref{sec:performance})}. When implemented, the TOTP derivation function need not re-calculate offset values corresponding to times in the future, and all offset calculation can be done in the background without blocking overall MFKDF key derivation or usage since $\sigma_F$ is output at step 2 of the above process.

\changes{
The proposed approaches for HOTP and TOTP-based key derivation require no modifications whatsoever to existing applications like Google Authenticator. Because $\mathit{hotpkey}_F$ is stored within $\alpha_F$ (encrypted as $ct_F$), the calculation of the next offset value(s) occurs entirely within the $\mathit{FactorSetup}$ and $\mathit{FactorUpdate}$ functions, and does not involve the authenticator application at all beyond the user obtaining $\mathit{otp}_{F,i}$ from the app once each login, as is always required.
} \\

\noindent \textbf{Theorem.} The HOTP and TOTP factor constructions yield computational ($d/log_{10}2$)-entropy factors.

\medskip

\noindent \textit{Proof.} Consider $F=(\sigma, \alpha)$ where $\sigma=\mathit{target}_F$ and $\alpha=(d,\mathit{ctr}_F,\mathit{offset}_F)$. Now $\sigma$ is uniformly random in $[0,10^d)$, and $d$, $\mathit{ctr}_F$, and $\mathit{offset}_F$ reveal no further information about $\sigma$, so for all $s$ and $a$ in support of $F$, the conditional probability $\mathit{Prob}(\sigma=s\mid\alpha=a)$ is at most $10^{-d}=2^{-(d/log_{10}2)}$. Thus, F is a statistical ($d/log_{10}2$)-entropy factor. Now consider $F'$ which adds $ct_F$ to $\alpha$; $F$ is computationally indistinguishable from $F'$ if $Enc$ is secure. $F'$ is the construction of \S\ref{sec:soft}. Thus the HOTP and TOTP factors are computational ($d/log_{10}2$)-entropy factors. \qed \newline

\noindent
\changes{
\textit{Remark.} Why should entropy be limited to $d/log_{10}2$ when HOTP/TOTP secrets are much larger than this? MFKDF cannot extract more entropy than the size of the \textit{witness} used to authenticate, regardless of the size of the underlying secret. In particular, because the user inputs a $d$-digit code to authenticate, MFKDF cannot possibly extract more entropy from this factor than the brute-force attack search space of all $10^d$ possible codes. Thus, the entropy gain from HOTP/TOTP factors in MFKDF is theoretically optimal with respect to the amount of entropy that is possible to derive from these factors.
}

\vspace{-1em}
\subsection{Hardware Tokens}
\label{sec:hard}
We next turn our attention to hardware authentication devices (``hard tokens'') like USB security keys and smart cards.
\changes{
While FIDO Universal 2nd Factor (U2F) and FIDO2 are the most commonly used standards for interacting with such devices, they are intentionally impossible to use for key derivation \cite{u2f}. Specifically, their inclusion of a hardware-generated random nonce in all signatures makes device responses completely nondeterministic even if the secret is known.
}
Fortunately, most hard tokens, including all YubiKeys \cite{yubikey}, also support authenticating via HMAC-SHA1 challenge-response, which can be used to derive an MFKDF factor.

In the simplest construction of this factor, the HMAC key $\mathit{hk}_F$ is itself used as $\sigma_F$. During a setup phase, a random challenge $c_{F,0}$ is generated, and the corresponding response $r_{F,0}$ is determined using $\mathit{hk}_F$. Both $c_{F,0}$ and ${\mathit{pad}_{F,0}=r_{F,0}\oplus\mathit{hk}_F}$ are stored in $\alpha_{F,0}$.\footnote{In this paper, $\oplus$ denotes bitwise XOR and $\odot$ denotes concatenation.} The factor derivation (with ${W_{F.i}=r_{F,i}}$) then proceeds as follows:

\begin{enumerate}[leftmargin=*]
    \itemsep -0.2em
    \item $c_{F,i}$ and $\mathit{pad}_{F,i}$ are extracted from $\alpha_{F,i}$.
    \item The HMAC key is recovered as $\mathit{hk}_F=W_{F,i}\oplus\mathit{pad}_{F,i}$.
    \item Generate a random 160-bit challenge $c_{F,i+1}$.
    \item Get the corresponding response using $\mathit{hk}_F$ with \mbox{HMAC-SHA1} ($\mathit{HS1}$): $r_{F,i+1}=\mathit{HS1}(\mathit{hk}_F,c_{F,i+1})$.
    \item Calculate the new pad, ${\mathit{pad}_{F,i+1}=r_{F,i+1}\oplus\mathit{hk}_F}$.
    \item Let ${\sigma_F=\mathit{hk}_F}$, ${\alpha_{F,i+1}=(c_{F,i+1},\mathit{pad}_{F,i+1})}$.
\end{enumerate}

\medskip

When a user wishes to sign in, they simply extract $c_{F,i}$ from $\alpha_{F,i}$, and generate $r_{F,i}$ with their YubiKey or equivalent device. This approach succeeds at generating static key material $\sigma_F$ from devices such as YubiKeys supporting HMAC-SHA1, while maintaining the freshness of non-repeating challenges and responses. An alternative approach could be to use a fixed random value for $\sigma_F$ and directly query the device with $c_{F,i+1}$ to get $\mathit{r}_{F,i+1}$, which has the benefit of not requiring $\mathit{hk}_F$ to be known to the KDF, but does not provide the same freshness.

\subsection{Out-of-band Authentication}
\label{sec:ooba}
\vspace{-0.4em}
Finally, we provide a general construction for out-of-band authentication (OOBA) factors such as SMS, email, and push notifications, which are currently amongst the most commonly used 2FA methods. Such factors, unlike all those previously described, are not completely trustless, instead requiring a degree of trust in the underlying channel (e.g., the cell carrier). The OOBA factor takes advantage of this by encrypting a dynamic OTP of $d$ digits under the public key of the channel ($\mathit{pk}_F$). The modular addition of said OTP with a fixed $\mathit{target_F}$ provides the same information-theoretic security as in the HOTP and TOTP constructions.

In the setup phase, a fixed $\mathit{target}_F$ and initial $\mathit{otp}_{F,0}$ are both chosen randomly in range $[0,10^d)$. As with HOTP/TOTP, the modular difference is stored as ${\mathit{offset}_{F,0}=(\mathit{target}_F-\mathit{otp}_{F,0})~\%~10^d}$. The parameters $\alpha_{F,0}$ consist of $(d,\mathit{pk}_F,\mathit{offset}_{F,0},\mathit{ct}_F)$, where $\mathit{ct}_F$ is now $\mathit{otp}_F$ encrypted under $\mathit{pk}_F$. The factor derivation process \mbox{(with $W_{F,i}=\mathit{otp}_{F,i}$)} then proceeds as follows:

\medskip

\begin{enumerate}[leftmargin=*]
    \itemsep -0.2em
    \item $(d,\mathit{pk}_F,\mathit{offset}_{F,i},\mathit{ct}_F)$ are recovered from $\alpha_{F,i}$.
    \item ${\sigma_F=\mathit{target}_F}$ is found as ${(\mathit{offset}_{F,i}+W_{F,i})~\%~10^d}$.
    \item The next OTP ($\mathit{otp}_{F,i+1}$) is chosen in range $[0,10^d)$.
    \item Calculate $\mathit{offset}_{F,i+1}$ as ${(\mathit{target}_F-\mathit{otp}_{F,i+1})~\%~10^d}$.
    \item Encrypt $\mathit{otp}_{F,i+1}$ under $\mathit{pk}_{F}$ as $\mathit{ct}_{F,i+1}$.
    \item Return $(d,\mathit{pk_{F}},\mathit{offset}_{F,i+1},\mathit{ct}_{F,i+1})$ as $\alpha_{F,i+1}$.
\end{enumerate}

\medskip

When a user wishes to authenticate via OOBA, they can obtain $\mathit{ct}_{F,i}$ from $\alpha_{F,i}$, submit $\mathit{ct}_{F,i}$ to the OOBA channel, and use $W_{F,i}=\mathit{otp}_{F,i}$ from the channel to derive $\sigma_F$. In practice, using the S/MIME key \cite{rfc3850} of the recipient as $\mathit{pk}_F$ is a good way to implement the OOBA factor for email, which can be extended to SMS using each cell carrier's \mbox{email-to-SMS} gateway service \cite{sms_gateway}.
\section{Entropy \& Brute-Force Resistance}
\label{sec:entropy}

\vspace{-0.5em}

\noindent Each of the factors presented in \S\ref{sec:factors} is a computational $m$-entropy factor with the value of $m$ shown in Tab. \ref{tab:entropy}.
\changes{For any given factor listed below, MFKDF extracts and fully utilizes the amount of entropy available in the witnesses used to authenticate. Thus, while it may be disappointing that any given factor does not provide more entropy, such limitations are inherently imposed by the factors themselves, not MFKDF.}

\begin{table}[H]
\resizebox{\columnwidth}{!}{%
\begin{tabular}{|cc|cc|}
\hline
\multicolumn{2}{|c|}{\multirow{2}{*}{\textbf{Factor}}} & \multicolumn{2}{c|}{\textbf{Entropy Bits ($m$ or $\lambda_F$)}} \\ \cline{3-4} 
\multicolumn{2}{|c|}{} & \multicolumn{1}{c|}{\textbf{General}} & \textbf{Typical} \\ \hline
\multicolumn{1}{|c|}{\multirow{4}{*}{\textbf{Constant}}} & Passwords & \multicolumn{1}{c|}{Varies} & $\approx40$ \cite{florencio_large_2006} \\ \cline{2-4} 
\multicolumn{1}{|c|}{} & Security Questions & \multicolumn{1}{c|}{Varies} & $\leq 10$ \cite{bonneau_secrets_2015} \\ \cline{2-4} 
\multicolumn{1}{|c|}{} & \begin{tabular}[c]{@{}c@{}}UUIDv4\\ (Recovery Code)\end{tabular} & \multicolumn{2}{c|}{122} \\ \hline
\multicolumn{1}{|c|}{\multirow{2}{*}{\textbf{\begin{tabular}[c]{@{}c@{}}Soft\\ Token\end{tabular}}}} & HOTP & \multicolumn{1}{c|}{$d/{\log_{10}2}$} & $\approx20$ when $d=6$ \\ \cline{2-4} 
\multicolumn{1}{|c|}{} & TOTP & \multicolumn{1}{c|}{$d/{\log_{10}2}$} & $\approx20$ when $d=6$ \\ \hline
\multicolumn{1}{|c|}{\textbf{\begin{tabular}[c]{@{}c@{}}Hard\\ Token\end{tabular}}} & \begin{tabular}[c]{@{}c@{}}HMAC-SHA1\\ (e.g., YubiKey)\end{tabular} & \multicolumn{2}{c|}{160} \\ \hline
\multicolumn{1}{|c|}{\textbf{OOBA}} & SMS, Email, \& Push & \multicolumn{1}{c|}{$d/{\log_{10}2}$} & $\approx20$ when $d=6$ \\ \hline
\end{tabular}%
}
\caption{Entropy of supported MFKDF factors.}
\label{tab:entropy}
\end{table}

A vitally important feature of the MFKDF construction in \S\ref{sec:mfkdf} is the application of a computationally-difficult KDF only after all factor material as been combined into $\mathit{mat}_K$. As a result, attackers cannot separately guess individual MFKDF factors, and must simultaneously correctly guess all factors to derive a key. Due to the ``avalanche effect'' (correlation freeness) \cite{al-kuwari_cryptographic_nodate} of the underlying KDF, if the derived key is incorrect, an attacker cannot easily determine which factor(s) were wrongly guessed. Therefore, given $n$ available factors with a mean factor entropy of $\bar{m}$ bits, the crack time $t$ will be $t\propto2^{\bar{m}}$ for PBKDFs and $t\propto2^{\bar{m} n}$ for MFKDF. We thus claim that $n$-factor MFKDF provides an exponential security improvement over PBKDFs; in other words, an MFKDF defined with respect to $n$ $m$-entropy factors $(F_1,F_2,...,F_n)$ is $nm$-entropy secure (see App. \ref{sec:proofs}). 

To illustrate the practical effect of this property, consider a typical PBKDF-derived key with $40$ bits of entropy and a 2-factor password-plus-HOTP MFKDF-derived key with $60$ bits of entropy. Assuming PBKDF2-SHA256-100,000 as the underlying PBKDF (the configuration used by LastPass~\cite{lastpass_security}), a 1 TH/s attacker\footnote{A single AntMiner S19 provides over 100 TH/s of SHA-256 \cite{s19pro}.} should require $\approx1.27$ days to crack the PBKDF key and $\approx3,653$ years to crack the MFKDF key, while the derivation time for the real user remains constant.
\section{Authenticating with Derived Keys}
\label{sec:auth}

\vspace{-0.5em}

So far, we have presented an MFKDF construction and a series of factor constructions that succeed at converting all of a user's authentication factors into a static key that can be used to encrypt user secrets. Doing so marks a paradigm shift in multi-factor authentication from software-based assurance to direct cryptographic protection of sensitive user data using all available authentication factors. The addition of secondary factors such as TOTP into the key derivation process allows user data to potentially remain secure under total system compromise, even in light of threats like credential stuffing. However, we pause for a moment to step back and consider the original purpose of these secondary factors: securely authenticating users.

A typical way of validating secondary factors like TOTP would be to store a user's TOTP key at a central authentication server. During authentication, the server generates the current TOTP code and compares it to the user-submitted value. Doing so, however, would largely defeat the purpose of using a TOTP factor in MFKDF, as an attacker who compromises the authentication server could steal the user's TOTP key and derive at least that portion of the user's MFKDF key. The same challenge exists with many of the factors presented in \S\ref{sec:factors}. Therefore, when using MFKDF, we require a means of verifying a user's factors that does not require server-side storage of factor-related secrets.

Thankfully, the security properties of MFKDF allow the derived key to itself be used to authenticate end users and implicitly verify all of their authentication factors. Because a properly-configured MFKDF key ($K$) cannot feasibly be derived without the presentation of all constituent factors, verifying a user's derivation of $K$ effectively constitutes verification of all factors. A standard key-based authentication algorithm like ISO/IEC 9798-2 Unilateral Authentication \cite{ISO97982} can therefore now be used to authenticate users.\footnote{The asymmetric variants of ISO 9798-2 can also be deployed here by using a symmetric MFKDF-derived key as a fixed seed to deterministically generate an asymmetric key pair via HMAC-DRBG \cite{barker_recommendation_2015}.}

\begin{figure}[h]
\includegraphics[width=0.75 \linewidth]{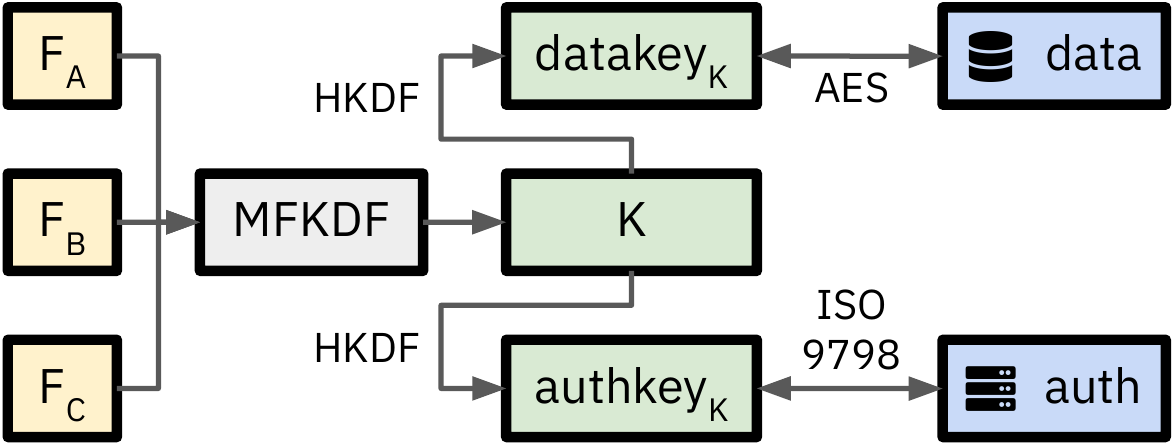}
\centering
\caption{MFKDF key establishment for authentication and confidentiality.}
\vspace{-1em}
\label{fig:auth}
\end{figure}

One final obstacle is that an MFKDF-derived key $K$ should not directly be shared with an authentication server if it is also to be relied on for data confidentiality. Thus, after $K$ is derived, separate sub-keys $\mathit{datakey}_K$ and $\mathit{authkey}_K$ should be derived using HKDF \cite{rfc5869}; $\mathit{datakey}_K$ can be used to encrypt user data (e.g., using AES \cite{fips197}) and $\mathit{authkey}_K$ can be used for authentication as described above. This approach, illustrated in Fig. \ref{fig:auth}, results in trustless cryptographic assurance of both data confidentiality and user authentication when using MFKDF. It is significantly advantageous over password-based authentication because a breach of the authentication server does not compromise confidentiality due to the use of separate client-side keys for encryption and authentication.

\section{Threshold MFKDF Construction}
\label{sec:threshold}

In \S\ref{sec:background}, we introduced the challenge of account recovery when using password-derived keys. This challenge exists, perhaps to an even greater extent, when using multi-factor derived keys, as the loss of any authentication factor could now imply total data loss if no further provisions are made. While said provisions typically entail using a master key with password-based key derivation, we describe in this section an equivalent recovery mechanism for MFKDF that does not implicate a central point of failure.

In a highly typical 2FA-enabled system, a user establishes a password, a secondary factor (e.g., HOTP), and a recovery factor (e.g., a recovery code). The password and secondary factor are used during normal logins; if the password is forgotten, the secondary factor and recovery factor can be used together to recover it, and similarly, if the secondary factor is forgotten, the password and recovery factor can be used together to recover it.

We note that the above policy is effectively equivalent to saying any $2$ of the $3$ established factors $\{\mathit{F_A},\mathit{F_B},\mathit{F_C}\}$ can be used to access an account, where in that case, $\mathit{F_A}$ is a password, $\mathit{F_B}$ is a HOTP code, and $\mathit{F_C}$ is a recovery code. We make explicit this concept of $2$-of-$3$ factors being required by introducing a general $t$-of-$n$ threshold MFKDF construction. In doing so, we allow MFKDF keys to be recovered on the client side as normal even if $n-t$ factors are lost, providing resilience against partial factor loss while codifying what is anyway the de facto standard for recovery.

\subsection{Threshold Multi-Factor Key Derivation}
\label{sec:tmfkdf}

As with the basic $n$-of-$n$ MFKDF construction in \S\ref{sec:mfkdf}, our threshold construction consists of separate setup and derive functions. The setup function uses $n$ factors to output a key of $\ell$ bits ($K\in\{0,1\}^\ell$) and $t$-of-$n$ parameters ($\alpha_{K,0}$), and the derive function uses at least $t$ of the $n$ factors and $\alpha_{K,i}$ to output $K$ and $\alpha_{K,i+1}$.

Let $\mathit{HKDF}$ be HKDF per RFC 5869 \cite{rfc5869}, $(\mathit{Share},\mathit{Comb},\mathit{Rec})$ be Shamir's secret sharing \cite{sss}, and $\mathit{KDF}$ be a hard PBKDF like Argon2 \cite{argon2}. Given $n$ factors ($\{F_1,F_2,...,F_n\}$), a threshold $t$, and security parameter $\ell$, a $k$-of-$n$ MFKDF key can be setup like so:

\begin{enumerate}[leftmargin=*]
    \itemsep -0.2em
    \item Sample $\sigma_K$ uniformly randomly in $\{0,1\}^{\ell}$.
    \item Split $\sigma_K$ into $n$ shares using $\mathit{Share}(\sigma_K,t,n)$.
    \item For each factor $F_j$, get factor material $\sigma_{F_j}$.
    \item Expand each $\sigma_{F_j}$ to size $\ell$ using HKDF.
    \item Combine each expanded factor with its corresponding share $\mathit{share}_{F_j}$ ($\mathit{pad}_{F_j}=\mathit{Enc}(\mathit{share}_{F_j},\sigma_{F_j})$) to get $\mathit{pads}_K$.
    \item Use $\mathit{KDF}$ on $\sigma_K$ to produce $K$.
    \item Get $\mathit{params}_{K,0}$ as $\alpha_{F,0}$ for each factor using $K$.
    \item Return $(t,\ell,\mathit{params}_{K,0},\mathit{pads}_{K})$ as $\alpha_{K,0}$.
\end{enumerate}

Using $t$ of the $n$ original factors ($\{F_1,F_2,...,F_t\}$) and $\alpha_{K,i}$, the derivation function is as follows:

\begin{enumerate}[leftmargin=*]
    \itemsep -0.2em
    \item Get $(t,\ell,\mathit{params}_{K,i},\mathit{pads}_{K})$ from $\alpha_{K,i}$.
    \item For each factor $F_j$, get factor material $\sigma_{F_j}$.
    \item Expand each $\sigma_{F_j}$ to size $\ell$ using HKDF.
    \item Decrypt each expanded factor with its corresponding pad in $\mathit{pads}_{K}$ to recover a share $\mathit{share}_{F_i}$.
    \item Combine shares using $Comb(\{\mathit{share}_{F_i},...,\mathit{share}_{F_k}\})$ to recover $\sigma_K$.
    \item Use $\mathit{KDF}$ on $\sigma_K$ to produce $K$.
    \item Get $\alpha_{F,i+1}$ using $K$ for each provided factor and update $\mathit{params}_{K,i}$ accordingly to produce $\mathit{params}_{K,i+1}$.
    \item Return $(t,\ell,\mathit{params}_{K,i+1},\mathit{pads}_{K})$ as $\alpha_{K,i+1}$.
\end{enumerate}

This threshold MFKDF construction takes advantage of the same feedback mechanism as \S\ref{sec:mfkdf} to facilitate dynamic OTP factors, and the factor constructions from \S\ref{sec:factors} can remain unchanged due to the overall modular design. Some aspects of the description are simplified for clarity; e.g., it omits the use of a salt. As before, a more complete formal statement of the scheme is given in \S\ref{sec:threshold_mfkdf_construction}.

\subsection{Recovery \& Reconstitution}
\label{sec:recovery}

Consider the $2$-of-$3$ key described above based on a password ($\mathit{F_A}$), HOTP code ($\mathit{F_B}$), and recovery code ($\mathit{F_C}$). Using the above $t$-of-$n$ threshold MFKDF construction with $t=2$, a user who has forgotten their password will still be able to derive their key using $\mathit{F_B}$ and $\mathit{F_C}$ and recover their account. This client-side recovery process constitutes a major security improvement by eliminating central master keys. However, the user may now want to replace $\mathit{F_A}$ with a new password $\mathit{F_{A_2}}$ moving forward, ideally without having to establish a brand new key and re-encrypt all secrets. The above $t$-of-$n$ threshold MFKDF construction is specifically designed to support this replacement operation like so:

\begin{enumerate}[leftmargin=*]
    \itemsep -0.2em
    \item Extract $(t,\ell,\mathit{params}_{K,in},\mathit{pads}_{K,in})$ from $\alpha_{K,in}$.
    \item Derive $\sigma_K$ from $t$ known factors as in \S\ref{sec:tmfkdf}.
    \item Get $\sigma_{\mathit{F_{A_2}}}$ and  $\alpha_{\mathit{F_{A_2}}}$ of the new factor. Update $\mathit{params}_{K,in}$ to reflect $\alpha_{\mathit{F_{A_2}}}$, yielding $\mathit{params}_{K,out}$.
    \item Recover $\mathit{share}_{\mathit{F_A}}$ ($j$th) as $\mathit{Rec}(\sigma_K,t,n,j)$.
    \item Update $\mathit{pad}_{\mathit{F_A}}$, e.g.,  $\mathit{pad}_{\mathit{F_{A_2}}}=\mathit{share}_{\mathit{F_A}}\oplus\sigma_{\mathit{F_{A_2}}}$.
    \item Update $\mathit{pads}_{K,in}$ to reflect $\mathit{pad}_{\mathit{FA}_2}$, yielding $\mathit{pads}_{K,out}$.
    \item Store $\alpha_{K,out}=(t,\ell,\mathit{params}_{K,out},\mathit{pads}_{K,out})$.
\end{enumerate}

The $\alpha_{K,out}$ produced by this operation is effectively identical to $\alpha_{K,in}$, other than the forgotten factor $\mathit{F_A}$ being replaced by $\mathit{F_{A_2}}$; the resulting key $K$ is the same. This process, termed ``key reconstitution,'' allows the factors constituting a threshold MFKDF-derived key to be updated while keeping the resulting $K$ static so that encrypted secrets remain accessible. The process can be extended to add or remove factors, or update the threshold $t$, without changing $K$ as long as $\sigma_K$ is known.
\section{Cryptographic Policy Enforcement}
\label{sec:policy}

The $t$-of-$n$ threshold MFKDF construction of \S\ref{sec:threshold} is sufficient for enabling many useful authentication schemes, but falls short of enforcing every combination of factors we might reasonably hope to enforce. For example, in a four factor setup with a password, HOTP code, recovery code, and security questions, a $2$-of-$4$ threshold MFKDF setup would enable recovering either the password or HOTP code using either the recovery code or security questions, but would also allow the key to be derived with just the recovery code and security questions. Furthermore, one may wish to assert that the password only be recovered using security questions, and that the HOTP factor only be recovered using a recovery code. To achieve this level of specificity in enforcement of exactly which factor combinations should be allowed to derive  a key requires a more granular key policy framework.

\subsection{Policy Framework}
\label{sec:policy_lang}

Consider a set of available authentication factors (${S=\{F_A,F_B,...\}}$). An allowable factor combination $C_i$ is defined as a non-empty subset of factors in S (${C_i \subseteq S, C_i \neq \varnothing}$) that can be used to derive a key ($\mathit{key}_K$). A policy $P$ is a set of all unique allowable factor combinations ($P=\{C_1,C_2,...\}$) to derive a key. A policy $P$ is said to be cryptographically enforced by a KDF if a user presenting a factor combination (${C=\{F_A,F_B,...\}}$) will be able to derive $\mathit{key}_K$ if and only if $C \in P$. Given a set of factors $S$, a fully expressive policy framework will be able to enforce any policy $P \in \mathcal{P}(\mathcal{P}(S) \setminus \varnothing) \setminus \varnothing$. For example, a fully-expressive policy framework for three authentication factors $(\mathit{F_A},\mathit{F_B},\mathit{F_C})$ will be able to enforce a policy $P$ defined by any non-empty subset of $\{\{\mathit{F_A}\},\allowbreak\{\mathit{F_B}\},\allowbreak\{\mathit{F_C}\},\allowbreak\{\mathit{F_A},\mathit{F_B}\},\allowbreak\{\mathit{F_B},\mathit{F_C}\},\allowbreak\{\mathit{F_A},\mathit{F_C}\},\allowbreak\{\mathit{F_A},\mathit{F_B},\mathit{F_C}\}\}$. 

To note explicitly a few non-requirements, asserting the absence of a factor (e.g., ``user does not know password $\mathit{F}_A$'') is not meaningful in this context. Allowing derivation based on an empty set of factors ($P_K=\{\varnothing\}$) is trivial but not particularly useful, and an empty policy ($P_K=\varnothing$), implying a key is impossible to derive, is also not useful.

Using this definition of fully-expressive policy enforcement, we present in this section a framework for enforcing arbitrarily complex authentication policies using MFKDF.

\subsection{Key Stacking}
\label{sec:stacking}
We introduce the notion of ``key stacking'' as a building block towards cryptographic policy enforcement. Key stacking entails using one multi-factor derived key as factor material for another multi-factor derived key. Therefore, when one wishes to derive an MFKDF key with stacked key factors, they may first need to derive one or more intermediate MFKDF keys, which are used as inputs to the final MFKDF key derivation. The many intentional symmetries in the $\mathit{Setup}$ and $\mathit{Derive}$ functions of MFKDF and those of MFKDF factors, namely the use of a $\alpha_F$ or $\alpha_K$, and output of fixed $\sigma_F$ or $\sigma_K$, may provide sufficient intuition into the key stacking method, but a formal construction is still given in \S\ref{sec:factor_constructions} for completeness.

To illustrate how key stacking enables enforcement of previously impossible factor combinations, consider again the four-factor setup described above with a password ($F_A$), HOTP code ($F_B$), recovery code ($F_C$), and security questions ($F_D$), whereby the password should only be recovered using security questions, and the HOTP should factor only be recovered using a recovery code. With key stacking, we can now enforce this pattern like so:

\begin{enumerate}[leftmargin=*]
    \itemsep -0.2em
    \item Let $K_P$ be a $1$-of-$2$ MFKDF key using $\{F_A,F_B\}$.
    \item Let $K_Q$ be a $1$-of-$2$ MFKDF key using $\{F_A,F_C\}$.
    \item Let $K_R$ be a $1$-of-$2$ MFKDF key using $\{F_D,F_B\}$.
    \item Via key stacking, $F_P=K_P$, $F_Q=K_Q$, $F_R=K_R$.
    \item Let $K$ be a $1$-of-$3$ MFKDF key using $\{F_P,F_Q,F_R\}$.
\end{enumerate}

Per the above construction, $K$ can be derived using a password and HOTP code, password and recovery code, or security questions and HOTP code, but not using security questions and recovery code, password and security questions, or HOTP and recovery code. Therefore, our desired authentication policy has been achieved. In \S\ref{sec:policy_crypto}, we will show that this technique is sufficient to enforce arbitrarily complex authentication policies.

When using key stacking, only the final MFKDF derivation needs a hard underlying PBKDF, and all intermediate MFKDF derivations can use a fast KDF like HKDF or even return $\sigma_K$ directly. We thus expect about $10$~ms of additional overhead per stacked key (see \S\ref{sec:performance}). For security reasons, only the final derived key should be fed back to the factor update functions for all constituent factors.

\subsection{Policy Enforcement}
\label{sec:policy_crypto}

Per the completeness definition of \S\ref{sec:policy_lang}, threshold MFKDF and key stacking are sufficient to cryptographically enforce any desired authentication policy $P=\{C_1,C_2,...\}$ given a set of factors ${S=\{F_A,F_B,...\}}$ as follows:

\begin{enumerate}[leftmargin=*]
    \itemsep -0.2em
    \item For all $C_i \in P$, let $K_i$ be an $n$-of-$n$ key with $\forall F \in C_i$.
    \item Let $\{F_{K_1},F_{K_2},...\}$ be $\{{K_1},{K_2},...\}$ via key stacking.
    \item Let $K$ be a $1$-of-$n$ key with $\{F_{K_1},F_{K_2},...\}$.
\end{enumerate}

Per the above construction, a user presenting a factor combination $C$ will be able to derive $K$ via a stacked sub-key $K_i$ if and only if ${C \in P}$; this is true for any ${K_i=P \in \mathcal{P}(\mathcal{P}(S) \setminus \varnothing) \setminus \varnothing}$.
Note while the above construction proves MFKDF can achieve arbitrary authentication policy enforcement, it is not necessarily the most efficient way to implement any given authentication policy. The example of \S\ref{sec:policy_example} illustrates a more direct and concretely efficient implementation of a particular policy.

\subsection{Security Statement}
\label{sec:policy_security}

\noindent \textbf{Theorem.} Let $\mathit{HKDF}$ be $(t,q,\varepsilon)$-secure. Given a set of independent computational $m_i$-entropy factors $S$ and any policy $P \in \mathcal{P}(\mathcal{P}(S) \setminus \varnothing) \setminus \varnothing$, let $E_i$ denote $\sum\{m_1,m_2,...,m_n\}$ for all ${C_i \in P}$. Let ${j=\mathit{argmin}_i E_i}$, and $F_P$ be a combination of $C_j$. Then the policy-based MFKDF construction of \S\ref{sec:policy_mfkdf_construction} w.r.t $\mathit{HKDF}$ and $P$ shall be \mbox{$(t,q,\varepsilon)$-secure} w.r.t. $F_P$.

\medskip

\noindent \textit{Proof Sketch.} Full proofs are given in Appendix \ref{sec:proofs}. We begin by showing that $n$ factors can be combined to produce a factor with entropy equal to the sum of input factor entropy. We use this to show that the basic MFKDF (\S\ref{sec:mfkdf}) is secure w.r.t. the combination of its factors. Next, we show that in the $1$-of-$n$ case, the threshold MFKDF (\S\ref{sec:tmfkdf}) is secure w.r.t. its lowest-entropy factor. Finally, because policy MFKDF (\S\ref{sec:policy_crypto}) is a composition of basic $n$-of-$n$ MFKDF and threshold $1$-of-$n$ MFKDF, it is secure w.r.t. the sum of factor entropy in the weakest allowable combination $C \in P$.

\subsection{Practical Example}
\label{sec:policy_example}

Consider the following nuanced but fairly realistic authentication policy: \newline

\begin{enumerate}[leftmargin=*]
    \itemsep -0.2em
    \item Users MUST authenticate using a password and TOTP.
    \item Users MAY recover either factor using security questions or a UUIDv4 recovery code.
    \item Users MUST also always use email OTP for recovery.
    \item Users MAY bypass TOTP when using a known device.\footnote{A device factor was not discussed in \S\ref{sec:factors}, but can be constructed quite easily by persisting factor material on a trusted device, e.g., as a cookie.} \newline
\end{enumerate}

\begin{figure}[h]
\includegraphics[width=\linewidth]{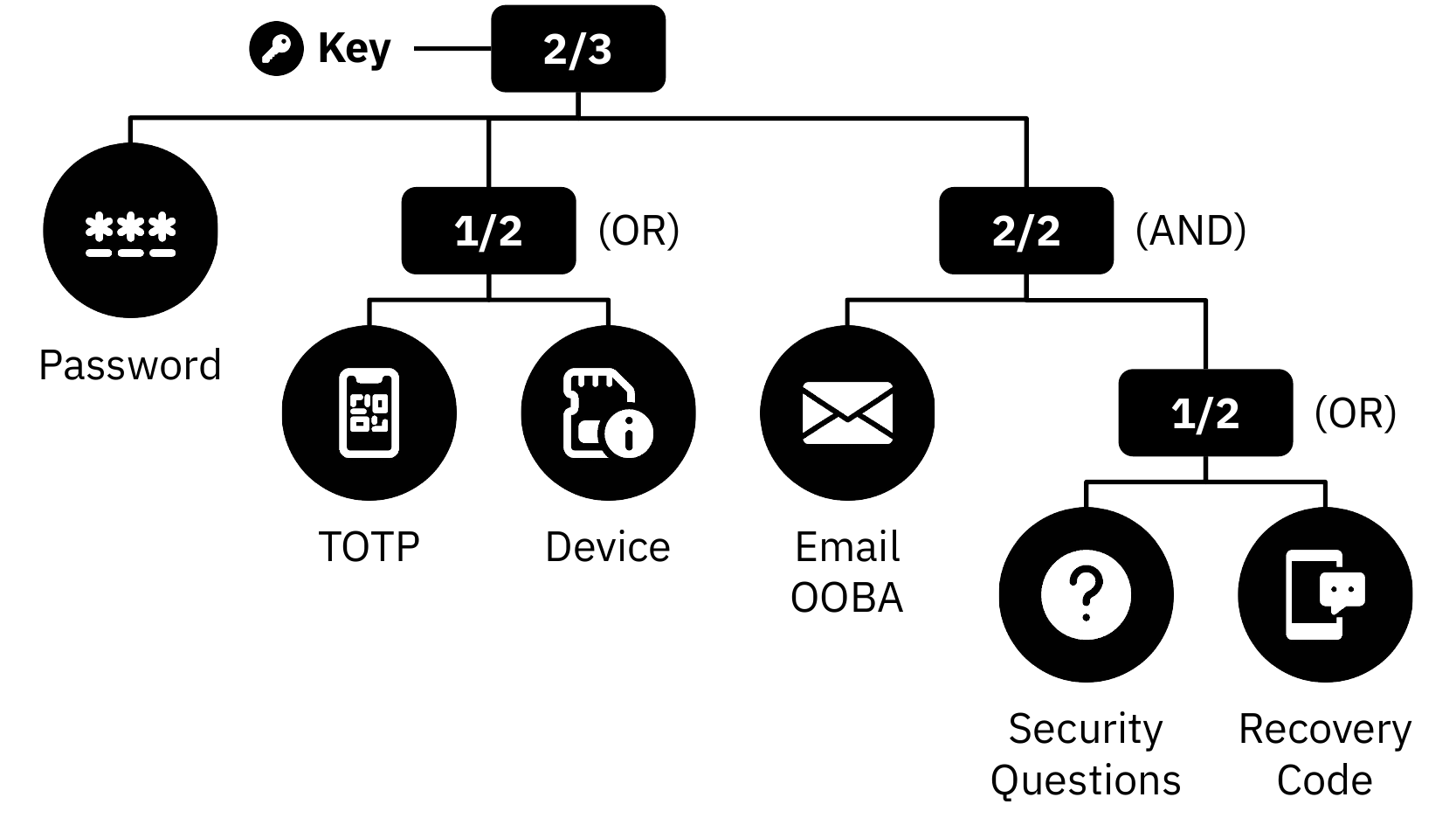}
\centering
\caption{MFKDF factor tree enforcing sample authentication policy.}
\label{fig:central_policy}
\end{figure}

The above policy can be enforced using threshold MFKDF with key stacking, as shown in Fig. \ref{fig:central_policy}. In fact, we implement (and cryptographically enforce) this exact policy in our centralized proof-of-concept system in \S\ref{sec:centralized}. 

\section{Applications}
\label{sec:applications}

To illustrate the immediate practical utility of MFKDF, we implemented and evaluated a fully-featured MFKDF library in JavaScript, which we used to produce two proof-of-concept applications. The first proof-of-concept (\S\ref{sec:centralized}) is a centralized full-stack web application used to demonstrate the use of MFKDF in current PBKDF applications. The second proof-of-concept (\S\ref{sec:decentralized}) is a fully decentralized application intended to illustrate the use of MFKDF in environments where PBKDFs would not be viable. We also used this library to perform a performance evaluation in \S\ref{sec:performance}.

\vspace{-0.3em}

\subsection{Centralized Proof-of-Concept}
\label{sec:centralized}

\vspace{-0.5em}

In \S\ref{sec:background}, we motivated the need for MFKDF by using password managers as an archetypal example of PBKDF-based systems. We now revisit the example of password management by discussing our implementation of a secure MFKDF-based password management application. To this end, we implemented and evaluated a full-stack password management web application using MFKDF for both \mbox{authentication} and encryption of stored passwords. All MFKDF operations take place on the client side; the host can be fully untrusted. \changes{The integrity of the client-side JavaScript code, including the MFKDF library, must be ensured, such as through the use of Subresource Integrity (SRI) tags or a trusted CDN. This is exactly the deployment model currently used by major PBKDF2-based web applications such as LastPass and Dashlane, with PBKDF2 simply being replaced by MFKDF in our application.}

Using the method of \S\ref{sec:policy}, an MFKDF-based password management application can implement and cryptographically enforce an authentication policy of its choosing; in our proof-of-concept application, we chose to use the policy of \S\ref{sec:policy_example}. Per this policy, a user can log in normally using a password and TOTP code, and can recover either factor if lost by using email OOBA along with either a recovery code or security questions. When enforced with MFKDF, this policy represents a $10^6$-times increase in brute-force difficulty over PBKDFs, and is a major improvement over existing password management systems which use a central master key to facilitate recovery per SP 800-57 \cite{sp800_57}. We also implemented the recovery and reconstitution method of \S\ref{sec:recovery}, allowing users to change a forgotten factor after recovery without re-encrypting secrets.

\begin{figure}[H]
    \centering
    \begin{subfigure}[b]{0.323\linewidth}
        \centering
        \includegraphics[width=\textwidth]{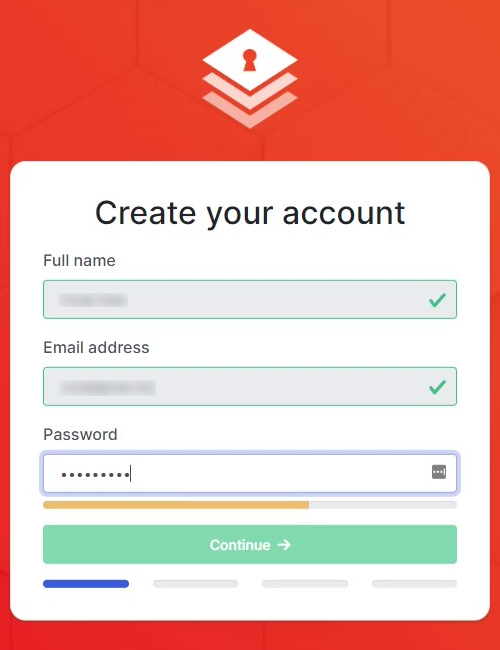}
        \caption{Signup}
        \label{fig:signup}
    \end{subfigure}
    \hfill
    \begin{subfigure}[b]{0.323\linewidth}
        \centering
        \includegraphics[width=\textwidth]{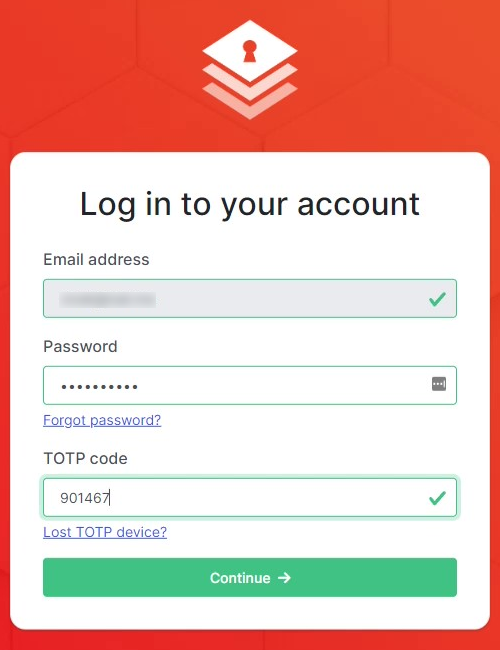}
        \caption{Login}
        \label{fig:login}
    \end{subfigure}
    \hfill
    \begin{subfigure}[b]{0.323\linewidth}
        \centering
        \includegraphics[width=\textwidth]{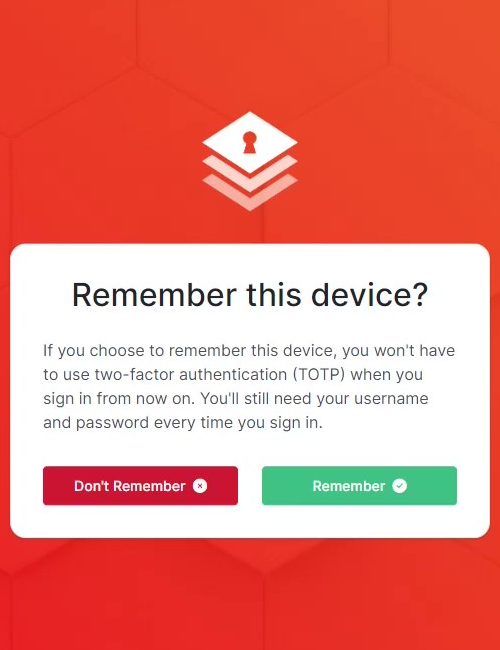}
        \caption{Persistence}
        \label{fig:persistence}
    \end{subfigure}
    
    \vspace{0.04 \linewidth}
    \begin{subfigure}[b]{0.655\linewidth}
        \centering
        \includegraphics[width=\textwidth]{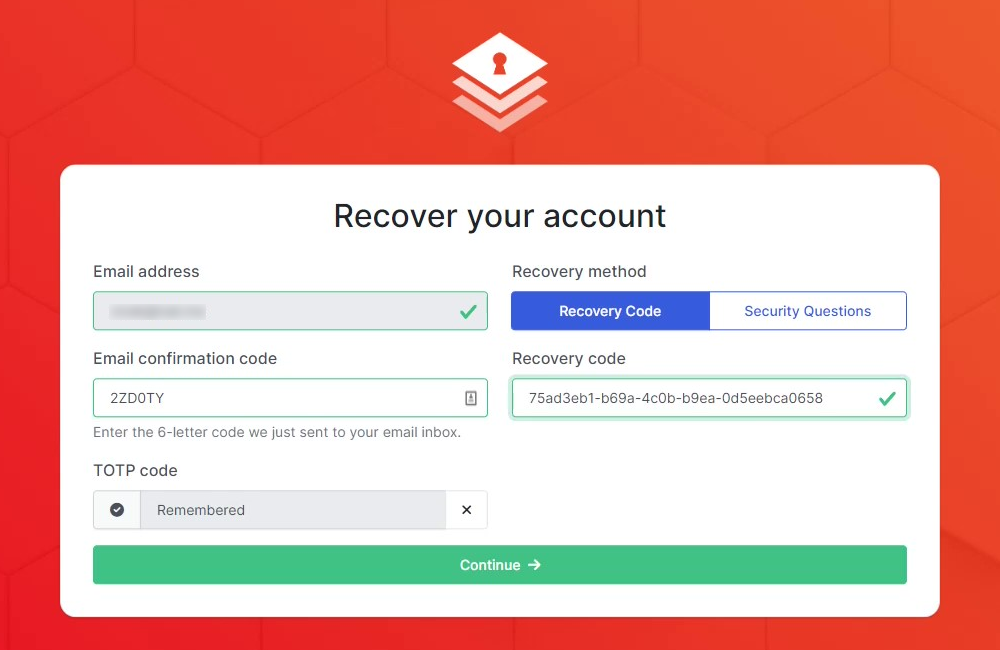}
        \caption{Recovery}
        \label{fig:recovery}
    \end{subfigure}
    \hfill
    \begin{subfigure}[b]{0.33\linewidth}
        \centering
        \includegraphics[width=\textwidth]{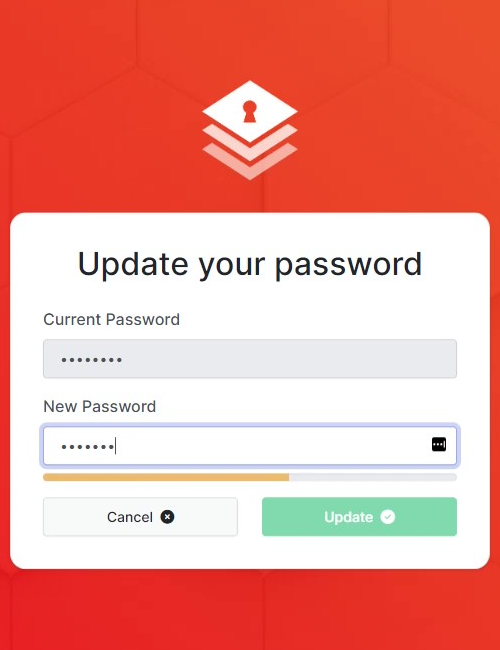}
        \caption{Reconstitution}
        \label{fig:reconstitution}
    \end{subfigure}
    
    \caption{Authentication screens for centralized demo application.}
    \label{fig:centralized}
\end{figure}

Aside from the various security advantages of using MFKDF in this application, our major takeaway from this endeavor is with respect to the usability of MFKDF. MFKDF does not demand the use of new authentication factors purpose-built for key derivation, instead supporting the same (unmodified) factors like HOTP, TOTP, YubiKey, or OOBA that users are familiar with and likely already using \changes{(e.g., we verified that our application was fully compatible with the latest version of Google Authenticator.)} They can therefore use the same signup, login, and recovery processes, with MFKDF operating transparently in the background to provide enhanced security with no tangible impact on the UI or UX (see Fig. \ref{fig:centralized}). We also showed that the implementation of common convenience features like skipping MFA on trusted devices and recovering and reconstituting lost factors are not hindered by the use of MFKDF. The use of MFKDF during the login and setup processes introduced less than $100$~ms of overhead to each.

\subsection{Decentralized Proof-of-Concept}
\label{sec:decentralized}
To demonstrate the potential use of MFKDF in fully decentralized applications, we implemented a distributed cryptocurrency wallet that supports ``logging in'' with traditional authentication factors (username, password, and MFA), but relies on no committees or trust assumptions. In essence, the wallet key is derived directly from the user's authentication factors with MFKDF whenever it is needed, rather than storing the key (or shares thereof) on any device. While the idea of such a wallet has already been proposed \cite{cryptoeprint:2021/1522}, its implementation hasn't been possible without MFKDF. 

\begin{figure}[H]
\includegraphics[width=\linewidth]{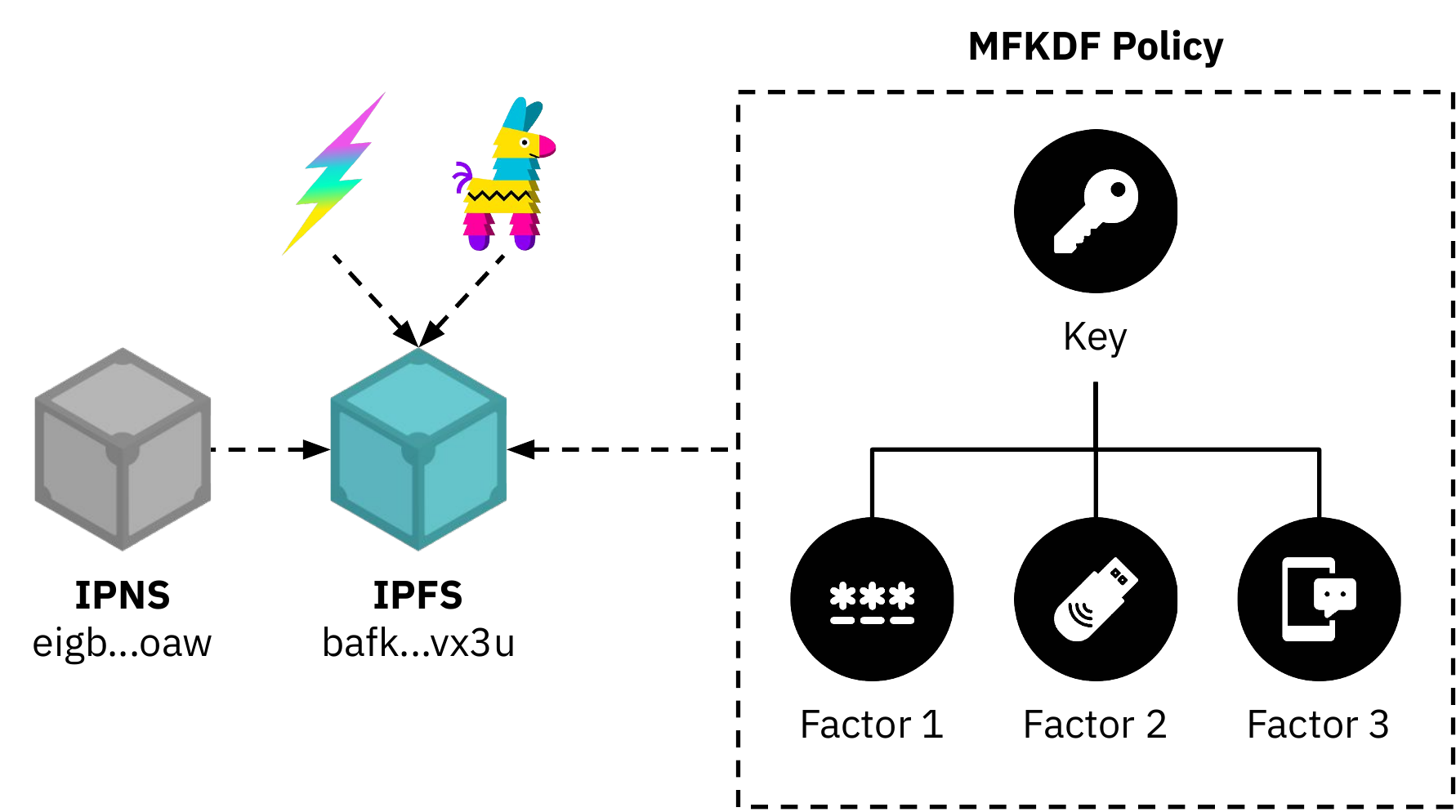}
\centering
\caption{Architecture of decentralized MFKDF wallet demo application}
\label{fig:decentralized}
\end{figure}

Fig. \ref{fig:decentralized} illustrates the architecture we used to implement the proof-of-concept decentralized MFKDF wallet. Users create their wallet by establishing a 2-of-3 threshold MFKDF key. Upon creation, the policy document is uploaded to the InterPlanetary File System (IPFS) \cite{benet_ipfs_2014} and a corresponding InterPlanetary Name System (IPNS) \cite{ipns} record is created, the address of which becomes the ``username.''\footnote{In the future, Ethereum Name Service (ENS) \cite{ens} could be used instead of IPNS to facilitate human-readable usernames.} A user ``logs in'' to their wallet using said username along with at least two of their three authentication factors, allowing them to retain access to their wallet even if any one factor is lost or forgotten. If the MFKDF policy document is updated, such as to reconstitute a lost factor or upon each login for factors like HOTP, the new policy is uploaded to IPFS and the IPNS record is updated accordingly.

% \begin{figure}[h]
% \includegraphics[width=\linewidth]{083_FIG_Wallet.png}
% \centering
% \caption{User interface of decentralized MFKDF wallet demo application}
% \label{fig:wallet}
% \end{figure}

Our MFKDF-based wallet has many of the traditional benefits of custodial wallets (namely, portability, recoverability, and multi-factor authentication with familiar authentication factors) while in fact being decentralized, trustless, and non-custodial.
We present this proof-of-concept mainly to emphasize that the potential applications for MFKDF reach far beyond the situations where PBKDFs would typically be deployed. PBKDFs would not be considered feasible for such an application due to the lower key entropy, potential for credential stuffing, password spraying, and brute-force, and the inability to recover from a lost factor.
\section{Performance Evaluation}
\label{sec:performance}
To evaluate the performance of MFKDF in a practical setting, we benchmarked a fully-featured JavaScript implementation of MFKDF in Chrome Browser v103.0.5060.114 on Windows 10 v21H2. Our test device used an AMD Ryzen 9 5950X (16-core, 3.4 GHz), although only single-thread performance is relevant in this browser setting.

\begin{figure}[h]
\includegraphics[width=\linewidth]{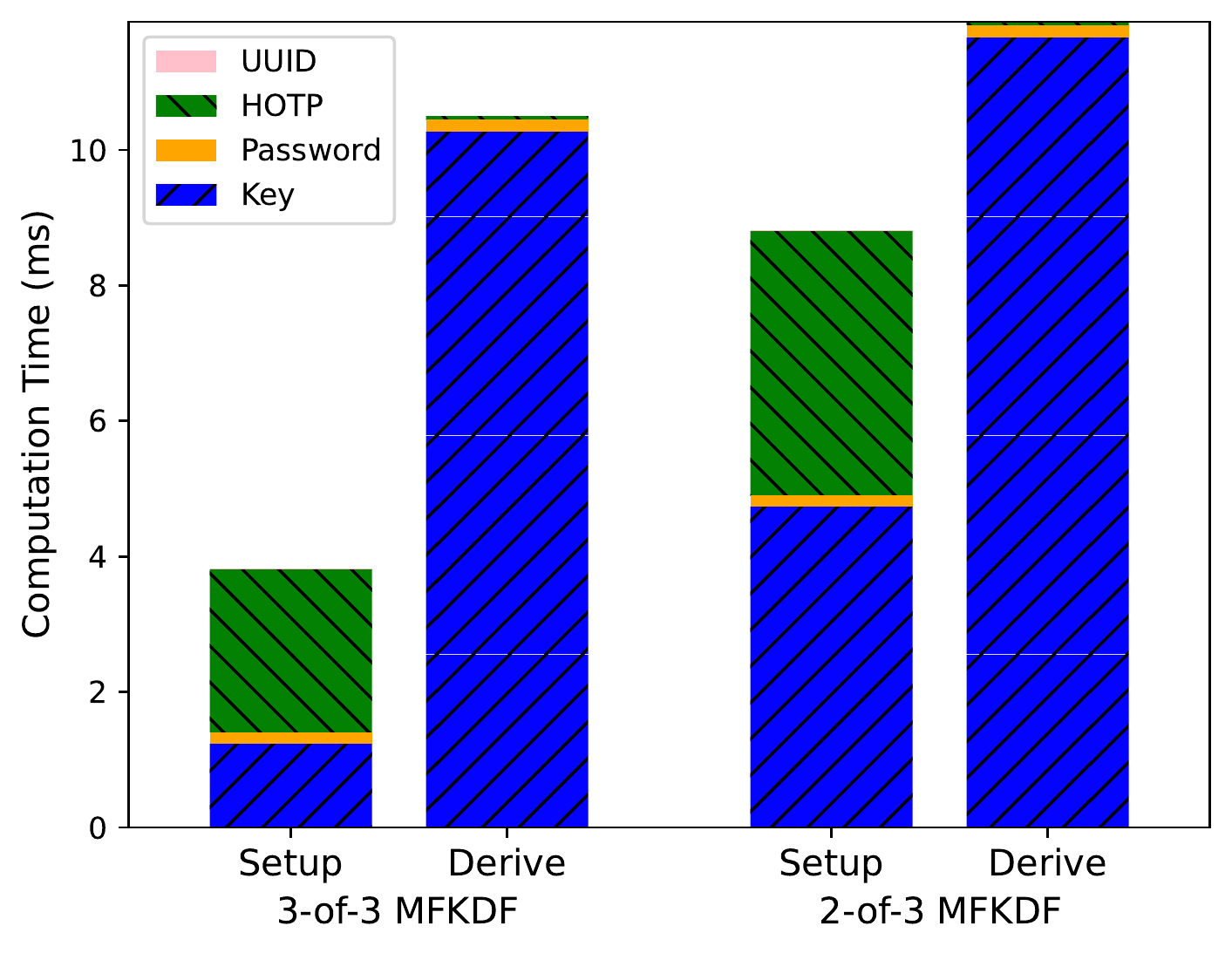}
\centering
\caption{Setup and derivation time of 2-of-3 and 3-of-3 MFKDF.}
\label{fig:mfkdf_perf}
\end{figure}

\subsection{MFKDF Performance}
Fig. \ref{fig:mfkdf_perf} shows the setup and derivation time for a \mbox{3-of-3} MFKDF key and \mbox{2-of-3} threshold MFKDF key based on the mean of $N=100$ setup and derive iterations with password, HOTP, and recovery code (UUIDv4) factors.

The 3-of-3 MFKDF setup and derivation had a mean computation time of $\bar{x} = 3.84$~ms and $\bar{x} = 10.52$~ms respectively. For the 2-of-3 threshold MFKDF, these were $\bar{x} = 8.83$~ms and $\bar{x} = 11.90$~ms. In both cases, HKDF was used as the KDF to isolate the overhead of MFKDF from the computational difficulty of the underlying KDF.

\subsection{Factor Performance}
\begin{figure}[h]
\includegraphics[width=\linewidth]{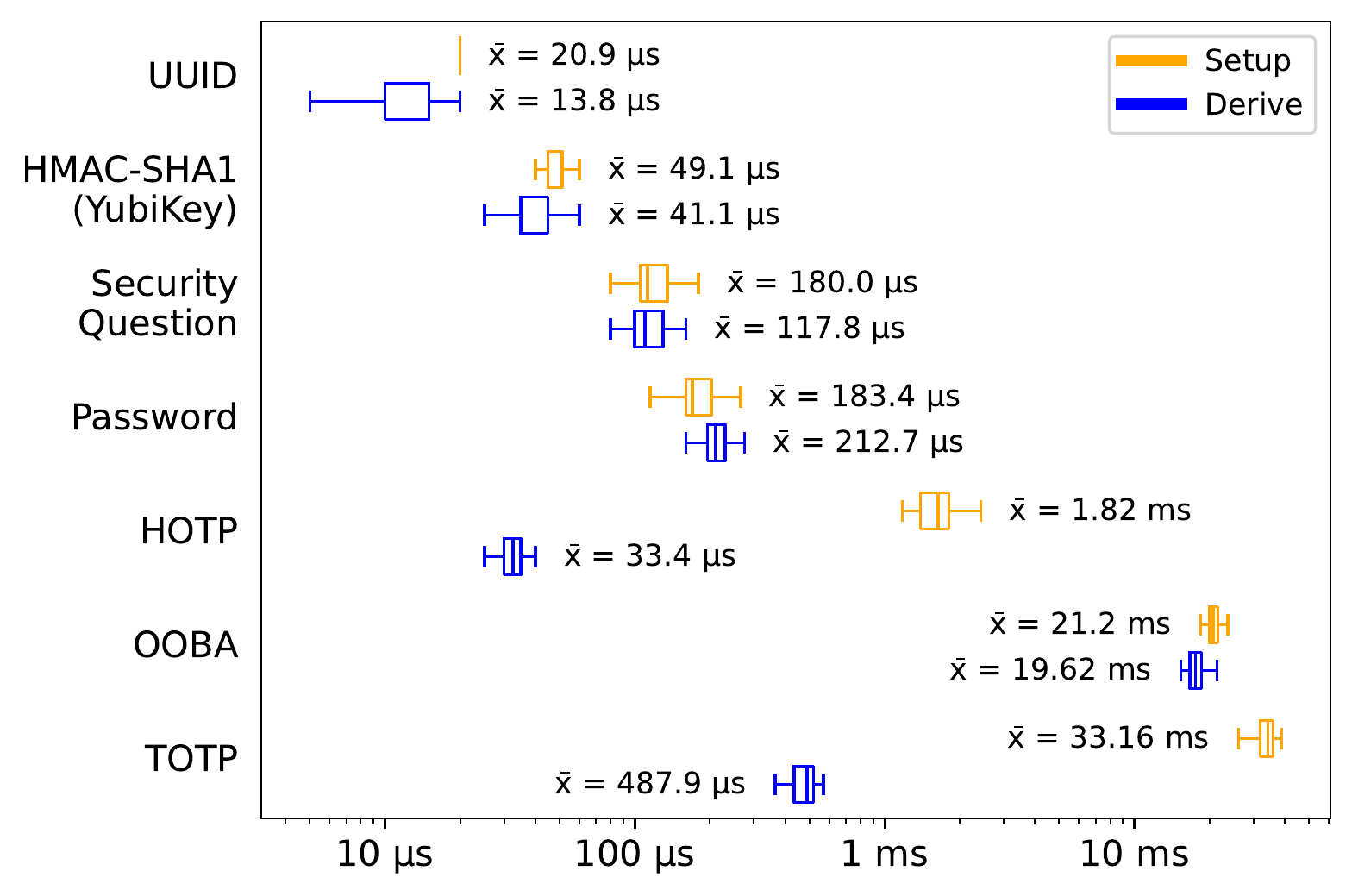}
\centering
\caption{Setup and derivation time of all supported MFKDF factors.}
\label{fig:factor_perf}
\end{figure}

Fig. \ref{fig:factor_perf} shows the range of computation times for the setup and derive functions of each supported factor across $100$ iterations. No individual factor has a setup or derive time of more than $2$~ms, other than OOBA ($\bar{x} = 21.2$~ms, $\bar{x} = 19.62$~ms) and TOTP ($\bar{x} = 33.16$~ms, $\bar{x} = 0.49$~ms).

The overhead of the TOTP factor is highly dependent on the time window parameter used. The above results reflect a window of $2920$~cycles ($24$~hours). If the time window is increased to $87600$~cycles ($30$~days), the setup time is $\bar{x} = 841$~ms and derivation time is $\bar{x} = 9.31$~ms. \\

\noindent \textbf{Discussion.} We noted in \S\ref{sec:background} that PBKDFs are intentionally computationally difficult so as to slow the rate of brute-force attacks. Most sources target a key derivation time of $250$~ms. While we used HKDF instead of a PBKDF in the above tests to isolate the overhead of the MFKDF construction from the intentional difficulty of the underlying KDF, the results should be understood in the context of this $250$~ms target for the underlying KDF. Thus, even the $20$~ms overhead of the slowest factors (other than TOTP with a large $w$) is negligible.
\vspace{-1em}
\section{Related Work}
\label{sec:related}
\vspace{-1em}
There are no known works that describe a general-purpose multi-factor key derivation function or key derivation using OTP authentication factors like TOTP and HOTP, based on a thorough search of terms such as ``multi-factor key derivation,'' ``two-factor key derivation,'' and ``hotp/totp key derivation.''

Several prior works have proposed two-factor key derivation approaches based specifically on a YubiKey hardware token and a password \cite{7414154, passwordscon2014}. \changes{Entirely biometric-based key derivation is also a widely studied problem \cite{biokey, uzun_cryptographic_2021, seo_construction_2018, Soutar1998BiometricEU}.}

Many works propose alternatives to key derivation for solving the key management problem, in particular by secret sharing a key amongst several user devices or \cite{dalskov_2fe_2020} a trusted committee \cite{cryptoeprint:2016/144, cryptoeprint:2021/1522}. Such solutions require a majority of the committee or devices to be honest and uncompromised, an assumption we do not make in MFKDF.

Finally, several works address the related but distinct problem of multi-factor authenticated key exchange \cite{pointcheval_multi-factor_2008, liu_multi-factor_2011, chen_modular_2014, 6920371}. Unlike key derivation, authenticated key exchange focuses on establishing a secure communication channel between two trusted parties in the presence of man-in-the-middle adversaries. Thus, multi-factor authenticated key exchange (MFAKE) is an effective replacement for password-authenticated key exchange (PAKE), but not for password-based key derivation functions (PBKDFs) as MFKDF is.
\vspace{-1em}

\section{Discussion}
\vspace{-1em}
While the concept of ``multi-factor key derivation'' may be apparent from the name itself, the combination of PBKDFs with MFA is non-trivial due to the dynamic nature of popular factors. In this paper, we overcome this challenge by converting dynamic factors into static keying material using data encrypted with the output key, essentially allowing secrets to be safely stored within the key derivation function itself. The key material for several factors can then be concatenated to generate a key via a standard KDF, or via secret sharing to produce flexible threshold and policy variants.

Despite this relatively simple structure, MFKDF has the potential to improve new and existing applications alike. Returning to our motivating example of password managers, these applications in particular stand to benefit, allowing secret keys to be derived from multiple authentication factors rather than just from passwords (or, in some cases, directly managed by users). Further, secure account recovery can now be facilitated even if a recovery key was not previously stored on a trusted user device. Moreover, the threat of credential stuffing is significantly reduced; while attackers require just a single attempt to check a compromised credential against a PBKDF-encrypted ciphertext, MFKDF leaves one or more entire factors intact which must be separately cracked.

\subsection{Limitations}
There are several distinct scenarios in which an MFKDF-derived key may be compromised, even if correctly implemented with secure primitives.
First, it is expected that an MFKDF-derived key is only as secure as its underlying factors, and that these factors must themselves be properly managed and protected. For example, if an SMS device is vulnerable to a SIM-swapping attack, then the corresponding MFKDF OOBA factor may also be at risk.

Further, if a combination of factors with insufficient entropy is chosen, brute-force attacks may still be feasible. As noted in \S\ref{sec:policy}, MFKDF-derived keys inherit the entropy of the weakest combination of factors allowed to derive them.

What is perhaps less obvious is that a combination of these two attacks may also be feasible. For example, while an MFKDF key combining a password and OTP factor may itself be infeasible to crack, the compromise of the password factor may allow the remaining factor to be defeated by brute-force.

Finally, the key could be compromised if it is not properly managed after derivation (e.g., via spyware or a browser vulnerability), as is the case with PBKDF-derived keys.

\subsection{Future Work}
\vspace{-0.5em}
\noindent \textbf{Applications.} In \S\ref{sec:background}, we described the wide variety of systems currently depending on PBKDFs. Many of these systems can be enhanced by the use of MFKDF. We hope to perhaps see corporate networking protocols encrypting traffic using all authentication factors, or enterprise operating systems supporting disk encryption based on MFKDF. Of course, many password managers or cloud storage services would benefit greatly from MFKDF. As illustrated in \S\ref{sec:decentralized}, MFKDF also has potential utility in decentralized applications. Overall, we look forward to seeing many systems, new and existing, either enhanced or made possible by MFKDF.

\smallskip

\noindent \textbf{Factors.} While the constructions given in \S\ref{sec:factors} account for the vast majority of current MFA usage \cite{soar}, there remain important authentication methods for which factor constructions are not currently known. Chief among these are intrinsic factors like biometrics, geolocation, device fingerprinting, and behavioral authentication. Deriving key material based on single sign-on, in particular via OIDC, would also be very helpful. We hope to see specific constructions for these, as well as ``general'' factor constructions, perhaps based on MPC or trusted hardware, for arbitrary factors.

\smallskip

\noindent \textbf{Extensions.} Further, we suggest a number of theoretical extensions to MFKDF that would further extend its utility. First, while it was not our focus here, we suggest investigating the use of MFKDF as a simple replacement to password hashing. Next, we suggest extensions improving the forward-security of MFKDF. We also suggest integrating MFKDF with encrypted databases to facilitate useful applications on MFKDF-encrypted data, including the sharing of encrypted objects between multiple users. Lastly, we hope to see works illustrating the backwards-compatibility of MFKDF with unmodified PBKDF-based systems, such as via browser extensions.

\smallskip

\noindent \textbf{Usability.} Finally, while many aspects of our design were motivated by usability, no user study has yet been performed. Future work should therefore focus on empirically comparing the usability of MFKDF with that of PBKDFs.

\vspace{-0.5em}
\section{Conclusion}
\label{sec:conclusion}
\vspace{-0.6em}

PBKDFs continue to play an outsized role in widely-used applied cryptographic systems compared to the research attention they receive. MFKDF offers a multi-axis improvement over PBKDFs, providing better security and additional functionality with no changes to the authentication factors nor noticeable impact on the UI or UX of the end user. It represents a fast, flexible, secure, and practical solution to the key management problem, with an emphasis on solving issues like factor recovery, authentication policy enforcement, plug-and-play compatibility with existing PBKDF-based systems, and support for existing widely-used authentication factors. In doing so, it changes the status quo of MFA from software verification to direct cryptographic assurance.

\section*{Acknowledgements}
We thank Guru Vamsi Policharla, Deevashwer Rathee, Xiaoyuan Liu, Gonzalo Munilla Garrido, Julien Piet, Sriram Sridhar, Jens Ernstberger, and Sanjam Garg for their advice and feedback. We also thank our anonymous shepherd for their guidance.
This work was supported in part by the National Science Foundation, by the National Physical Science Consortium, and by the Fannie and John Hertz Foundation.
Any opinions, findings, and conclusions or recommendations expressed in this material are those of the authors and do not necessarily reflect the views of the supporting entities.

\section*{Availability}
The source code for a JavaScript library implementing all of the MFKDF features described in this paper is available at:

\begin{itemize}[itemsep=0em]
    \item \href{https://github.com/multifactor/mfkdf}{github.com/multifactor/mfkdf}
\end{itemize}

\noindent We invite interested parties to visit \href{https://mfkdf.com}{mfkdf.com} for detailed documentation, tutorials, unit tests, and code coverage reports. The centralized demo, decentralized demo, and benchmarking scripts used in this paper are available at these repositories:

\begin{itemize}[itemsep=0em]
    \item \href{https://github.com/multifactor/mfkdf-application-demo}{github.com/multifactor/mfkdf-application-demo}
    \item \href{https://github.com/multifactor/mfkdf-wallet-demo}{github.com/multifactor/mfkdf-wallet-demo}
    \item \href{https://github.com/multifactor/mfkdf-benchmark}{github.com/multifactor/mfkdf-benchmark}
\end{itemize}

{
\small
\bibliographystyle{plainurl}
\bibliography{999_REFS.bib}
}

\appendix

\clearpage

\section{Algorithms}
\label{sec:algorithms}

\vspace{-0.6em}

\subsection{Formalization}
\label{sec:formalization}

\vspace{-0.6em}

Let an MFKDF $\mathit{FactorInstance}$ $F_i$ be a tuple ${(\sigma_F\in\{0,1\}^{\ell_F},(\mathit{key}\mapsto\alpha_{F,i}))}$ representing the $i$th derivation of a given factor, where $\sigma_F$ is the factor-specific key material (static), $\ell_F$ is the number of bits of entropy in $F$, and ${(key\mapsto\alpha_{F,i})}$ is a function that takes the final derived $\mathit{key}$ and returns the factor policy ($\alpha_{F,i}$), a tuple whose contents vary by factor.

An MFKDF factor construction consists of a $\mathit{FactorSetup}$ function, which instantiates a factor given a configuration $\mathit{cfg}_F$ (which varies by factor), and a $\mathit{FactorDerive}$ function, which derives a factor from an instance-specific ${W_{F,i}\in\{0,1\}^{\ell_F}}$ and a $\alpha_{F,i}$ output by $\mathit{FactorSetup}$ or a previous derivation: \newline

\noindent ${\mathit{FactorSetup}:}$

${\mathit{cfg}_F\mapsto\mathit{FactorInstance}~F_0}$

\noindent ${\mathit{FactorDerive}:}$ 

${W_{F,i},\alpha_{F,i}\mapsto\mathit{FactorInstance}~F_{i+1}}$ \newline

Similarly, let an MFKDF $\mathit{KeyInstance}$ $K_i$ be a tuple ${(\alpha_{K,i}, \mathit{key}\in\{0,1\}^{\ell_K})}$ representing the $i$th derivation of a key, where $\mathit{key}$ is the derived key (static), $\ell_K$ is the size of the derived key in bits, and $\alpha_{K,i}$ varies.

An MFKDF construction consists of a $\mathit{KeySetup}$ function, which instantiates a key of size $\ell_K$ given an array of $\mathit{FactorInstance}$s, and a $\mathit{KeyDerive}$ function, which derives an established key from an array of $\mathit{FactorInstance}$s and a $\alpha_{K,i}$: \newline

\noindent ${\mathit{KeySetup}:}$

${\ell_K,\mathit{FactorInstance}[]~\mathit{FS}_0\mapsto\mathit{KeyInstance}~K_0}$

\noindent ${\mathit{KeyDerive}:}$

${\mathit{FactorInstance}[]~\mathit{FS}_i,\alpha_{K,i}\mapsto\mathit{KeyInstance}~K_{i+1}}$ \newline

In \S\ref{sec:factor_constructions}, we will provide MFKDF factor constructions for several popular authentication factors. In \S\ref{sec:mfkdf_construction}, we will provide a basic MFKDF construction, and in \S\ref{sec:threshold_mfkdf_construction}, we will provide a threshold MFKDF construction. We use $\odot$ for bitstring and array concatenation, and $\oplus$ for bitwise XOR.

\hfill\null

\vspace{-0.6em}

\subsection{Factor Constructions}
\label{sec:factor_constructions}

\vspace{-0.6em}

\begin{algorithm}[H]
\caption{Factor Construction for Constant Factors}
\label{alg:mfkdf_constant_factor}
\begin{algorithmic}[1]
\Function{Setup}{$\mathit{cfg}_F: ({W_{F}\in\{0,1\}^{\ell_F}})$}
    \State $\sigma_F \gets \mathit{transform}(W_{F})$
    \State \Return $(\mathit{key} \rightarrow (),\sigma_F)$
\EndFunction
\Function{Derive}{${W_{F}\in\{0,1\}^{\ell_F}}, \alpha_{F,i}$}
    \State $\sigma_F \gets \mathit{transform}(W_{F})$
    \State \Return $(\mathit{key} \rightarrow (),\sigma_F)$
\EndFunction
\end{algorithmic}
\end{algorithm}

We begin with the factor construction for constant factors like passwords, security questions, and recovery codes, shown in algorithm \ref{alg:mfkdf_constant_factor}. This is by far the simplest factor construction, with just an optional $\mathit{transform}$ step between the input $W$ and output $\sigma_F$. The $\mathit{transform}$ step can be used, for example, to standardize the case of a security answer, or to extract useful data from a UUIDv4 recovery code.

\vspace{-1em}

\begin{algorithm}[H]
\caption{Factor Construction for HMAC-SHA1}
\label{alg:mfkdf_hmacsha1_factor}
\begin{algorithmic}[1]
\Require Let $\mathit{HS1}$ be HMAC-SHA1 per RFC 2014 \cite{rfc2104}.
\Function{Setup}{$\mathit{cfg}_F: ({\mathit{hmackey}_{F}\in\{0,1\}^{160}})$}
    \State $\mathit{challenge}_{F,0} \gets \{0,1\}^{160}$
    \State $W_{F,0} \gets \mathit{HS1}(\mathit{hmackey}_{F}, \mathit{challenge}_{F,0})$
    \State $\mathit{pad}_{F,0} \gets W_{F,0}\oplus\mathit{hmackey}_{F}$
    \State $\alpha_{F,0} \gets (\mathit{challenge}_{F,0},\mathit{pad}_{F,0})$
    \State \Return $(\mathit{key} \rightarrow \alpha_{F,0},\mathit{hmackey}_F)$
\EndFunction
\Function{Derive}{${W_{F,i}\in\{0,1\}^{160}}, \alpha_{F,i}$}
    \State $(\mathit{challenge}_{F,i},\mathit{pad}_{F,i}) \gets \alpha_{F,i}$
    \State $\mathit{hmackey}_F \gets W_{F,i}\oplus\mathit{pad}_{F,i}$
    \State $\mathit{challenge}_{F,i+1} \gets \{0,1\}^{160}$
    \State ${\mathit{response}_{F,i+1} \gets \mathit{HS1}(\mathit{hmackey}_{F}, \mathit{challenge}_{F,i+1})}$
    \State $\mathit{pad}_{F,i+1} \gets \mathit{challenge}_{F,i+1}\oplus\mathit{hmackey}_{F}$
    \State $\alpha_{F,i+1} \gets (\mathit{challenge}_{F,i+1},\mathit{pad}_{F,i+1})$
    \State \Return $(\mathit{key} \rightarrow \alpha_{F,i+1},\mathit{hmackey}_F)$
\EndFunction
\end{algorithmic}
\end{algorithm}

\vspace{-1em}

Algorithm \ref{alg:mfkdf_hmacsha1_factor} shows the MFKDF factor construction for HMAC-SHA1 challenge-response, an authentication method implemented by many hardware tokens such as \mbox{YubiKeys}. Since the $\mathit{challenge}_{F_i}$ and the corresponding response ($\mathit{response}_{F,i+1}$) are in effect uniformly random and non-repeating, they can be used as a one-time pad for the $\mathit{hmackey}_{F}$ (which is itself the $\sigma_{F}$) with information theoretic security. The HMAC-SHA1 factor fixes $\ell_F=160$ due to the output of SHA1 being exactly 20 bytes.

\vspace{-1em}

\begin{algorithm}[H]
\caption{Factor Construction for OOBA}
\label{alg:mfkdf_ooba_factor}
\begin{algorithmic}[1]
\Require Let $(\mathit{Enc},\mathit{Dec})$ be public-key encryption.
\Function{Setup}{$\mathit{cfg}_F: (d,\mathit{pk}_F)$}
    \State $\mathit{target}_F \gets \mathbb{N}\cup[0,10^d)$
    \State $\mathit{otp}_{F,0} \gets \mathbb{N}\cup[0,10^d)$
    \State $\mathit{offset}_{F,0} \gets (\mathit{target}_F - \mathit{otp}_{F,0})~\%~10^{d}$
    \State $\mathit{ct}_{F,0} \gets \mathit{Enc}(\mathit{otp}_{F,0}, \mathit{pk}_F)$
    \State \Return $(\mathit{key} \rightarrow (d,\mathit{pk}_F,\mathit{ct}_{F,0},\mathit{offset}_{F,0}),\mathit{target}_F)$
\EndFunction
\Function{Derive}{${W_{F,i}\in\mathbb{N}\cup[0,10^d)}, \alpha_{F,i}$}
    \State $(d,\mathit{pk}_F,\mathit{ct}_{F,i},\mathit{offset}_{F,i}) \gets \alpha_{F,i}$
    \State $\mathit{target}_{F} \gets (\mathit{offset}_{F,i} + W_{F,i})~\%~10^d$
    
    \State $\mathit{otp}_{F,i+1} \gets \mathbb{N}\cup[0,10^d)$
    \State $\mathit{offset}_{F,i+1} \gets (\mathit{target}_F - \mathit{otp}_{F,i+1})~\%~10^{d}$
    \State $\mathit{ct}_{F,i+1} \gets \mathit{Enc}(\mathit{otp}_{F,i+1}, \mathit{pk}_F)$
    
    \State $\alpha_{F,i+1} \gets (d,\mathit{pk}_F,\mathit{ct}_{F,i+1},\mathit{offset}_{F,i+1})$
    \State \Return $(\mathit{key} \rightarrow \alpha_{F,i+1},\mathit{target}_F)$
\EndFunction
\end{algorithmic}
\end{algorithm}

Algorithm \ref{alg:mfkdf_ooba_factor} shows the MFKDF factor construction for out-of-band authentication (OOBA) factors such as email or SMS. Unlike all other factors presented herein, OOBA factors are not perfectly trustless; they inherently depend on the honesty of the underlying channel (e.g., email provider, phone carrier). The OOBA factor takes advantage of this by pre-encrypting a numeric OTP of $d$ digits using the public key of the channel. The modular addition of the OTP with a fixed $\mathit{target_F}$ provides the same information theoretic security as the one-time pad for the HMAC-SHA1 factor.

\vspace{-1em}

\begin{algorithm}[H]
\caption{Factor Construction for HOTP}
\label{alg:mfkdf_hotp_factor}
\begin{algorithmic}[1]
\Require Let $\mathit{HOTP}$ be HOTP per RFC 4226 \cite{rfc4226}.
\Require Let $(\mathit{Enc},\mathit{Dec})$ be symmetric-key encryption.
\Function{Setup}{$\mathit{cfg}_F: (d, {\mathit{hotpkey}_{F}\in\{0,1\}^{\ell_F}})$}
    \State $\mathit{target}_F \gets \mathbb{N}\cup[0,10^d)$
    \State $\mathit{otp}_{F,0} \gets HOTP(\mathit{hotpkey}_{F},1)~\%~10^{d}$
    \State $\mathit{offset}_{F,0} \gets (\mathit{target}_F - \mathit{otp}_{F,0})~\%~10^{d}$
    \Function{Policy}{$key$}
        \State $\mathit{ct}_F \gets Enc(\mathit{hotpkey}_{F},\mathit{key}) $
        \State \Return $(d,1,\mathit{offset}_{F,0},\mathit{ct}_F)$
    \EndFunction
    \State \Return $(\textsc{Policy},\mathit{target}_F)$
\EndFunction
\Function{Derive}{${W_{F,i}\in\mathbb{N}\cup[0,10^d)}, \alpha_{F,i}$}
    \State $(d,\mathit{ctr}_{F,i},\mathit{offset}_{F,i},\mathit{ct}_F) \gets \alpha_{F,i}$
    \State $\mathit{target}_{F} \gets (\mathit{offset}_{F,i} + W_{F,i})~\%~10^d$
    \State $\mathit{ctr}_{F,i+1} \gets \mathit{ctr}_{F,i} + 1$
    \Function{Policy}{$key$}
        \State $\mathit{hotpkey}_{F} \gets Dec(\mathit{ct}_{F},\mathit{key}) $
        \State $\mathit{otp}_{F,i+1} \gets HOTP(\mathit{hotpkey}_{F},\mathit{ctr}_{F,i+1})~\%~10^{d}$
        \State $\mathit{offset}_{F,i+1} \gets (\mathit{target}_F - \mathit{otp}_{F,i+1})~\%~10^{d}$
        \State \Return $(d,\mathit{ctr}_{F,i+1},\mathit{offset}_{F,i+1},\mathit{ct}_F)$
    \EndFunction
    \State \Return $(\textsc{Policy},\mathit{target}_F)$
\EndFunction
\end{algorithmic}
\end{algorithm}

\vspace{-1em}

Algorithms \ref{alg:mfkdf_hotp_factor} and \ref{alg:mfkdf_totp_factor} show MFKDF factor constructions for HOTP and TOTP respectively, each supporting a variable number of OTP digits $d$ (typically, $d=6$).
HOTP is the first factor to take advantage of the key feedback mechanism described in \S\ref{sec:mfkdf}, allowing the $\mathit{hotpkey}_F$ to be securely embedded within $\alpha_F$ and used to set up $F_{i+1}$ during the derivation of $F_{i}$ by incrementing $ctr_{F,i}$. The TOTP and HOTP constructions takes advantage of the same information theoretic blinding via modular arithmetic as is used in the OOBA factor.
The TOTP construction is similar to HOTP, with the addition of a window parameter $w$. By default, we suggest $w=87600$ cycles (30 days). Note that by definition, ${\mathit{TOTP}(K)=\mathit{HOTP}(K,\lfloor(T-T_0)/T_X\rfloor)}$. Per our construction, the $\mathit{offset}$ value is calculated for every possible time between $T$ and $T+wT_X$; this does not reveal anything useful to attackers due to the forward security of TOTP.
For the OOBA, HOTP, and TOTP constructions presented above, $\ell_F=d/{\log_{10}2}$ ($\ell_F\approx20$ when $d=6$). 

\vspace{-1em}

\begin{algorithm}[H]
\caption{Factor Construction for TOTP}
\label{alg:mfkdf_totp_factor}
\begin{algorithmic}[1]
\Require Let $\mathit{HOTP}$ be HOTP per RFC 4226 \cite{rfc4226}.
\Require Let $T,T_0,T_X$ be TOTP times per RFC 6238 \cite{rfc6238}.
\Require Let $(\mathit{Enc},\mathit{Dec})$ be symmetric-key encryption.
\Function{Setup}{$\mathit{cfg}_F: (d,w, {\mathit{totpkey}_{F}\in\{0,1\}^{\ell_F}})$}
    \State $\mathit{target}_F \gets \mathbb{N}\cup[0,10^d)$
    \State $\mathit{ctr}_{F,0} \gets \lfloor(T-T_0)/T_X\rfloor$
    \For{$j \gets 0,w$}
        \State $\mathit{otp} \gets HOTP(\mathit{totpkey}_{F},\mathit{ctr}_{F,0}+j)~\%~10^{d}$
        \State $\mathit{offsets}_{F,0}[j] \gets (\mathit{target}_F - \mathit{otp})~\%~10^{d}$
    \EndFor
    \Function{Policy}{$key$}
        \State $\mathit{ct}_F \gets Enc(\mathit{totpkey}_{F},\mathit{key}) $
        \State \Return $(d,w,\mathit{ctr}_{F,0},\mathit{offsets}_{F,0},\mathit{ct}_F)$
    \EndFunction
    \State \Return $(\textsc{Policy},\mathit{target}_F)$
\EndFunction
\Function{Derive}{${W_{F,i}\in\mathbb{N}\cup[0,10^d)}, \alpha_{F,i}$}
    \State $(d,w,\mathit{ctr}_{F,i},\mathit{offsets}_{F,i},\mathit{ct}_F) \gets \alpha_{F,i}$
    \State $\mathit{ctr}_{F,i+1} \gets \lfloor(T-T_0)/T_X\rfloor$
    \State $\mathit{idx}_{F,i} \gets \mathit{ctr}_{F,i+1} - \mathit{ctr}_{F,i}$
    \State $\mathit{offset}_{F,i} \gets \mathit{offsets}_{F,i}[\mathit{idx}_{F,i}]$
    \State $\mathit{target}_{F} \gets (\mathit{offset}_{F,i} + W_{F,i})~\%~10^d$
    \Function{Policy}{$key$}
        \State $\mathit{hotpkey}_{F} \gets Dec(\mathit{ct}_{F},\mathit{key}) $
        \For{$j \gets 0,w$}
            \State ${\mathit{otp} \gets HOTP(\mathit{totpkey}_{F},\mathit{ctr}_{F,i+1}+j)~\%~10^{d}}$
            \State $\mathit{offsets}_{F,i+1}[j] \gets (\mathit{target}_F - \mathit{otp})~\%~10^{d}$
        \EndFor
        \State \Return $(d,w,\mathit{ctr}_{F,i+1},\mathit{offsets}_{F,i+1},\mathit{ct}_F)$
    \EndFunction
    \State \Return $(\textsc{Policy},\mathit{target}_F)$
\EndFunction
\end{algorithmic}
\end{algorithm}

\vspace{-1em}

We also include the factor construction for the ``stacked key'' factor in algorithm \ref{alg:mfkdf_stack_factor}, although its construction may already be evident from the symmetric definitions of an MFKDF construction and an MFKDF factor construction. The MFKDF input, $\mathit{FactorInstance}[]~\mathit{FS}$, is provided to the \textsc{Setup} and \mbox{\textsc{Derive}} functions as $W_F$; the $\mathit{KeySetup}$ or $\mathit{KeyDerive}$ function is invoked, and the derived key $\mathit{key_{K_F}}$ is returned as the factor key material $\sigma_F$. Thus, $\ell_F=\ell_{K_F}$ here.

\vspace{-1em}

\begin{algorithm}[H]
\caption{Factor Construction for Key Stacking}
\label{alg:mfkdf_stack_factor}
\begin{algorithmic}[1]
\Function{Setup}{$\mathit{cfg}_F: (\ell_K,\mathit{FactorInstance}[]~\mathit{FS})$}
    \State ${\mathit{KeyInstance}~K_{F,0} \gets \mathit{KeySetup}({\ell_K,\mathit{FS}})}$
    \State ${(\alpha_{K_F,0}, \mathit{key}_{K_F}}) \gets K_{F,0}$
    \State \Return $(\mathit{key} \rightarrow {\alpha_{K_F,0}},\mathit{key}_{K_F})$
\EndFunction
\Function{Derive}{${\mathit{FactorInstance}[]~\mathit{FS}_{F}}, \alpha_{F,i}$}
    \State ${\mathit{KeyInstance}~K_{F,i+1} \gets \mathit{KeyDerive}({\mathit{FS}_F,\alpha_{F,i}})}$
    \State ${(\alpha_{K_F,i+1}, \mathit{key}_{K_F}}) \gets K_{F,i+1}$
    \State \Return $(\mathit{key} \rightarrow \alpha_{K_F,i+1},\mathit{key}_{K_F})$
\EndFunction
\end{algorithmic}
\end{algorithm}

\subsection{MFKDF Construction}
\label{sec:mfkdf_construction}

\vspace{-0.6em}

Algorithm \ref{alg:mfkdf} is a simple $n$-of-$n$ MFKDF construction. Each authentication factor is converted into factor material $\sigma_F$ via its corresponding factor construction. The $\sigma_F$ for all factors are concatenated to form $\sigma_K$, which in turn is used to derive the $\mathit{key}$ via a PBKDF. The key feedback mechanism of \S\ref{sec:mfkdf} is then used to produce $\alpha_{K,i+1}$, making possible the use of dynamic factors like HOTP.

\vspace{-1em}

\begin{algorithm}[H]
\caption{MFKDF Construction}
\label{alg:mfkdf}
\begin{algorithmic}[1]
\Require Let $\mathit{KDF}$ be a hard PBKDF like Argon2 \cite{argon2}.
\Function{Setup}{${\ell_K,\mathit{FactorInstance}[]~\mathit{FS}_0}$}
    \State $\mathit{salt}_K \gets \{0,1\}^{\ell_K}$
    \State $\sigma_K \gets \varepsilon$
    \State $\mathit{fns}_{K,0} \gets []$
    \State $\mathit{fps}_{K,0} \gets []$
    \ForAll{$\mathit{FactorInstance}~F \in \mathit{FS}_0$}
        \State ${(\sigma_F,\mathit{fn}_F)} \gets F$
        \State $\sigma_K \gets \sigma_K \odot \sigma_F$ 
        \State $\mathit{fns}_{K,0} \gets \mathit{fns}_{K,0} \odot \mathit{fn}_F$ 
    \EndFor
    \State $\mathit{key} \gets \mathit{KDF}(\ell_K,\sigma_K,\mathit{salt}_K)$
    \ForAll{$\mathit{fn}_{K,0} \in \mathit{fns}_{K,0}$}
        \State $\alpha_{F,0} \gets \mathit{fn}_{K,0}(\mathit{key})$
        \State $\mathit{fps}_{K,0} \gets \mathit{fps}_{K,0} \odot \alpha_{F,0}$ 
    \EndFor
    \State \Return $((\ell_K,\mathit{salt}_K,\mathit{fps}_{K,0}),\mathit{key})$
\EndFunction
\Function{Derive}{$\mathit{FactorInstance}[]~\mathit{FS}_i,\alpha_{K,i}$}
    \State $(\ell_K,\mathit{salt}_K,\mathit{fps}_{K,i}) \gets \alpha_{K,i}$
    \State $\sigma_K \gets \varepsilon$
    \State $\mathit{fns}_{K,i} \gets []$
    \State $\mathit{fps}_{K,i+1} \gets []$
    \ForAll{$\mathit{FactorInstance}~F \in \mathit{FS}_i$}
        \State ${(\sigma_F,\mathit{fn}_F)} \gets F$
        \State $\sigma_K \gets \sigma_K \odot \sigma_F$ 
        \State $\mathit{fns}_{K,i} \gets \mathit{fns}_{K,i} \odot \mathit{fn}_F$ 
    \EndFor
    \State $\mathit{key} \gets \mathit{KDF}(\ell_K,\sigma_K,\mathit{salt}_K)$
    \ForAll{$\mathit{fn}_{K,i} \in \mathit{fns}_{K,i}$}
        \State $\alpha_{F,i+1} \gets \mathit{fn}_{K,i}(\mathit{key})$
        \State $\mathit{fps}_{K,i+1} \gets \mathit{fps}_{K,i+1} \odot \alpha_{F,i+1}$ 
    \EndFor
    \State \Return $((\ell_K,\mathit{salt}_K,\mathit{fps}_{K,i+1}),\mathit{key})$
\EndFunction
\end{algorithmic}
\end{algorithm}

\vspace{-3em}

\subsection{Threshold MFKDF Construction}
\label{sec:threshold_mfkdf_construction}

\vspace{-0.6em}

Algorithm \ref{alg:threshold_mfkdf} implements a $t$-of-$n$ threshold MFKDF. It expands upon the simple MFKDF construction by adding a threshold parameter $t$. The key material ($\sigma_K$) is split into $n$ shares via Shamir's secret sharing \cite{sss}, which are then padded by factor keys derived using HKDF \cite{rfc5869}. The one-time-pad can also be replaced by symmetric-key encryption as long as no checksums or integrity mechanisms are included in the scheme. Thus, if at least $t$ factors are provided, $t$ shares can be derived from their pads in $\alpha_K$ and thus $\sigma_K$ can be recovered.

\vspace{-1em}

\begin{algorithm}[H]
\caption{Threshold MFKDF Construction}
\label{alg:threshold_mfkdf}
\begin{algorithmic}[1]
\Require Let $\mathit{KDF}$ be a hard PBKDF like Argon2 \cite{argon2}.
\Require Let $\mathit{HKDF}$ be HKDF per RFC 5869 \cite{rfc5869}.
\Require \mbox{Let $(\mathit{Share},\mathit{Comb})$ be secret sharing (SSS) \cite{sss}.}
\Function{Setup}{${t,\ell_K,\mathit{FactorInstance}[]~\mathit{FS}_0}$}
    \State $\sigma_K \gets \{0,1\}^{\ell_K}$
    \State $\mathit{salt}_K \gets \{0,1\}^{\ell_K}$
    \State $\mathit{key} \gets \mathit{KDF}(\ell_K,\sigma_K,\mathit{salt}_K)$
    \State $\mathit{shares}_{K,0} \gets \mathit{Share}(\sigma_K,t,\mathit{len}(\mathit{FS}_0))$
    \State $\mathit{fps}_{K,0} \gets []$
    \State $\mathit{fs}_{K} \gets []$
    \ForAll{$\mathit{FactorInstance}~F \in \mathit{FS}_0$}
        \State ${(\sigma_F,\mathit{fn}_F)} \gets F$
        \State ${\mathit{pad}_F} \gets \mathit{HKDF}(\ell_K,\sigma_F,\varepsilon,\varepsilon)$
        \State ${\mathit{share}_F} \gets {\mathit{pad}_F}\oplus{\mathit{shares}_{K,0}[i]}$
        \State $\mathit{fs}_{K} \gets \mathit{fs}_{K} \odot \mathit{share}_F$
        \State $\alpha_{F,0} \gets \mathit{fn}_{F}(\mathit{key})$
        \State $\mathit{fps}_{K,0} \gets \mathit{fps}_{K,0} \odot \alpha_{F,0}$ 
    \EndFor
    \State \Return $((t,\ell_K,\mathit{salt}_K,\mathit{fps}_{K,0},\mathit{fs}_{K}),\mathit{key})$
\EndFunction
\Function{Derive}{$\mathit{FactorInstance}[]~\mathit{FS}_i,\alpha_{K,i}$}
    \State $(t,\ell_K,\mathit{salt}_K,\mathit{fps}_{K,i},\mathit{fs}_{K}) \gets \alpha_{K,i}$

    \State $\mathit{shares}_{K,i} \gets []$
    \State $\mathit{fns}_{K,i} \gets []$
    \State $\mathit{fps}_{K,i+1} \gets \mathit{fps}_{K,i}$
    \ForAll{$\mathit{FactorInstance}~F \in \mathit{FS}_i$}
        \State ${(\sigma_F,\mathit{fn}_F)} \gets F$
        \State ${\mathit{pad}_F} \gets \mathit{HKDF}(\ell_K,\sigma_F,\varepsilon,\varepsilon)$
        \State ${\mathit{share}_F} \gets {\mathit{pad}_F}\oplus{\mathit{fs}_{K}[j]}$
        \State $\mathit{shares}_{K,i} \gets \mathit{shares}_{K,i} \odot \mathit{share}_F$
        \State $\mathit{fns}_{K,i} \gets \mathit{fns}_{K,i} \odot \mathit{fn}_{F}$ 
    \EndFor
    \State $\sigma_K \gets \mathit{Comb}(\mathit{shares}_{K,i},t,\mathit{len}(\mathit{fs}_{K,i}))$
    \State $\mathit{key} \gets \mathit{KDF}(\ell_K,\sigma_K,\mathit{salt}_K)$
    \ForAll{$\mathit{fn}_{K,i} \in \mathit{fns}_{K,i}$}
        \State $\alpha_{F,i+1} \gets \mathit{fn}_{K,i}(\mathit{key})$
        \State $\mathit{fps}_{K,i+1}[j] \gets \alpha_{F,i+1}$ 
    \EndFor
    \State \Return $((t,\ell_K,\mathit{salt}_K,\mathit{fps}_{K,i+1},\mathit{fs}_{K}),\mathit{key})$
\EndFunction
\end{algorithmic}
\end{algorithm}

\vspace{-2em}

\subsection{Policy MFKDF Construction}
\label{sec:policy_mfkdf_construction}

\vspace{-2em}

\begin{algorithm}[H]
\caption{Policy MFKDF Construction}
\label{alg:policy_mfkdf}
\begin{algorithmic}[1]
\Function{Setup}{${t,\ell_K,\mathit{FactorInstance}[][]~\mathit{P}}$}
    \State $\mathit{FactorInstance[]}~R \gets []$
    \ForAll{$\mathit{FactorInstance[]}~C \in \mathit{P}$}
        \State $\mathit{SK} \gets \textsc{Alg\ref{alg:mfkdf_stack_factor}Setup}((\ell_K, C))$
        \State $R \gets R \odot SK$
    \EndFor
    \State \Return $\textsc{Alg\ref{alg:threshold_mfkdf}Setup}(1,\ell_K,R)$
\EndFunction
\end{algorithmic}
\end{algorithm}

% To summarize the method of \S\ref{sec:policy}, given a set of available authentication factors (${S=\{F_A,F_B,...\}}$), for any policy $P \in \mathcal{P}(\mathcal{P}(S) \setminus \varnothing) \setminus \varnothing$, a policy-based MFKDF can be constructed like so:

% //
% \\
% //

\clearpage

\section{Security Arguments}
\label{sec:proofs}

\noindent \textbf{\mbox{Preliminary 1}.} \\ If ${P(A \mid C)=P(A \mid C \land D)}$, ${P(B \mid D)=P(B \mid C \land D)}$, then ${P(A \land B \mid C \land D)=P(A \mid C) \cdot P(B \mid D)}$. \newline

\noindent \textbf{\mbox{Theorem 1.}} Let $F_a=(\sigma_a,\alpha_a)$ and $F_b=(\sigma_b,\alpha_b)$ be independent statistical $m_a$ and $m_b$-entropy factors. Let $F_a \times F_b$ denote $(\sigma_a \odot \sigma_b, \alpha_a \odot \alpha_b)$. Then $F_c=F_a \times F_b$ is a statistical \mbox{($m_a + m_b$)-entropy} factor. \\

\noindent \textit{Proof.} By Def. 2, for all $(s_a, a_a) \in F_a$ and $(s_b, a_b) \in F_b$, $P(\sigma_a=s_a\mid\alpha_a=a_a) \leq 2^{-m_a}$ and $P(\sigma_b=s_b\mid\alpha_b=a_b) \leq 2^{-m_b}$.
Now, for all ${(s_c, a_c) \in F_c}$, let $s_c=s_a \odot s_b$ and $a_c=a_a \odot a_b$, then $P(\sigma_c=s_c\mid\alpha_c=a_c) = P(\sigma_a=s_a \land \sigma_a=s_a \mid\alpha_a=a_a \land \alpha_b=a_b)$.
If $F_A$ and $F_B$ are independent, then ${P(\sigma_a=s_a \mid \alpha_a=a_a)=P(\sigma_a=s_a \mid \alpha_a=a_a \land \alpha_b=a_b)}$, ${P(\sigma_b=s_b \mid \alpha_b=a_b)=P(\sigma_b=s_b \mid \alpha_a=a_a \land \alpha_b=a_b)}$. 
Therefore, by Prelim. 1, $P(\sigma_c=s_c\mid\alpha_c=a_c)=P(\sigma_a=s_a \mid \alpha_b=a_b) \cdot P(\sigma_b=s_b\mid\alpha_b=a_b) \leq 2^{-m_a} \cdot 2^{-m_b}$.
Thus, $P(\sigma_c=s_c\mid\alpha_c=a_c) \leq 2^{-(m_a + m_b)}$, and $F_c=F_a \times F_b$ is a statistical \mbox{($m_a + m_b$)-entropy} factor. \qed \newline

\noindent \textbf{\mbox{Corollary 1.1.}}  Let $F_a$ and $F_b$ be independent computational $m_a$ and $m_b$-entropy factors. Then $F_c=F_a \times F_b$ is a computational \mbox{($m_a + m_b$)-entropy} factor. \\

\noindent \textit{Proof.} By Def. 3, there exist statistical $m_a$ and \mbox{$m_b$-entropy} factors $F_a'$ and $F_b'$ which are computationally indistinguishable from $F_a$ and $F_b$. Let $F_c' = F_a' \times F_b'$. By Thm. 1, $F_c'$ is a statistical \mbox{($m_a + m_b$)-entropy} factor. Assume, for the sake of contradiction, that $F_c$ is computationally distinguishable from $F_c'$. Therefore, there exists a bit $b \in \sigma_c$ that can be distinguished from $\sigma_c'$. However, $\sigma_c=\sigma_a \odot \sigma_b$, so there must exist a bit $b \in \sigma_a$ or $b \in \sigma_b$ that can be distinguished from $\sigma_a'$ or $\sigma_b'$, but then $F_a$ and $F_b$ are not computational $m_a$ and $m_b$-entropy factors. Thus, $F_c$ is computationally indistinguishable from $F_c'$, and $F_c$ is a computational \mbox{($m_a + m_b$)-entropy} factor. \qed \newline

% Assume for contradiction that F_C is not a computational m-entropy factor, that is, it can be computationally distinguished from any m-entropy factor.

\noindent \textbf{\mbox{Corollary 1.2.}} Let ${S=\{F_1,F_2,...,F_n\}}$ be $n$ independent \mbox{computational} ${\{m_1,m_2,...,m_n\}}$-entropy factors.  Let ${\prod S}$ denote $F_1 \times F_2 \times ... \times F_n$. Let ${F_T=\prod S}$. Then $F_T$ is a computational \mbox{($\sum \{m_1,m_2,...,m_n\}$)-entropy} factor. \\

\noindent \textit{Proof.} Per Cor. 1.1, $F_1 \times F_2$ has entropy $m_1 + m_2$, then $F_1 \times F_2 \times ... \times F_n$ has entropy $m_1 + m_2 + ... + m_n$. Thus, $\prod S$ is a computational \mbox{($\sum m_i$)-entropy} factor. \qed \newline

\noindent \textbf{\mbox{Theorem 2.}} Let $\mathit{HKDF}$ be $(t,q,\varepsilon)$ ${\sum m_i}$-entropy secure. Then the \mbox{$n$-of-$n$} MFKDF construction of \S\ref{sec:mfkdf_construction} w.r.t. $\mathit{HKDF}$ and $S$ is \mbox{$(t,q,\varepsilon)$-secure} w.r.t. $F_T$. \newline

\noindent \textit{Proof.} Recall that $\sigma_T \in F_T = \sigma_1 \odot \sigma_2 \odot ... \odot \sigma_n$. In the construction of \S\ref{sec:mfkdf_construction}, the MFKDF output $K=\mathit{HKDF}(\sigma_T \in F_T)$. Per Cor. 1.2, $F_T=\prod S$ is a \mbox{($\sum m_i$)-entropy} factor. Since by Def. 7 HKDF is $(t,q,\varepsilon)$-secure w.r.t. all \mbox{($\sum m_i$)-entropy} factors, the MFKDF construction is \mbox{$(t,q,\varepsilon)$-secure} w.r.t. $F_T$. \qed \newline

\noindent \textbf{\mbox{Corollary 2.1.}} Let ${\mathit{Stack}(S)=(\sigma,\alpha)}$, ${\prod S = (\sigma_S, \alpha_S)}$ where $\alpha = \alpha_S$ and $\sigma$ is the output $K$ of the \mbox{$n$-of-$n$} MFKDF construction of \S\ref{sec:mfkdf_construction} w.r.t. $\mathit{HKDF}$ and $S$. Let $F_K=\mathit{Stack}(S)$. Then $F_K$ is a computational \mbox{($\sum m_i$)-entropy} factor. \newline

\noindent \textit{Proof.} Per Thm. 2, the $n$-of-$n$ MFKDF w.r.t. S is $(t,q,\varepsilon)$-secure w.r.t. $F_T = \prod S$. By Def. 6, no $(t,q,\varepsilon)$-attacker can distinguish between $K=\mathit{MFKDF}(\sigma_T,\ell)$ and $\{0,1\}^\ell$. 
Per Cor. 1.2, $F_T$ is a computational \mbox{($\sum_{i=0}^n m_i$)-entropy} factor. Assume $\ell \geq m$. Thus ${F_K=(K,\sigma \in F_T)}$ is computationally indistinguishable from $(\{0,1\}^\ell, \sigma \in F_T)$, and is a computational \mbox{($\sum m_i$)-entropy} factor. \qed \newline

\noindent \textbf{\mbox{Theorem 3.}} Let ${R=\{F_1,F_2,...,F_n\}}$ be $n$ \mbox{computational} ${\{m_1,m_2,...,m_n\}}$-entropy factors. Let ${j=\mathit{argmin}_i m_i}$. Let $\mathit{Enc}$ be $(t,q,\varepsilon)$ CPA-secure symmetric encryption. Then a \mbox{$1$-of-$n$} MFKDF construction per \S\ref{sec:threshold_mfkdf_construction} w.r.t. $\mathit{HKDF}$, $\mathit{Enc}$, and $R$ is \mbox{$(t,q,\varepsilon)$-secure} w.r.t. $F_j$. \newline

\noindent \textit{Proof.} In the $1$-of-$n$ case of SSS, $share_i$ = $secret$ for all $n$ shares. Thus per \S\ref{sec:threshold_mfkdf_construction}, $ct_i = \mathit{Enc}(\mathit{mat}_K, \sigma_i)$ for all $\sigma_i \in F_i \in R$. Given that $K=\mathit{HKDF}(\mathit{mat}_K)$ is \mbox{$(t,q,\varepsilon)$-secure} w.r.t. $\mathit{mat}_K$, and $\mathit{mat}_K$ is encrypted with $\sigma_j$, an adversary that can obtain $\mathit{mat}_K$ without $\sigma_j \in F_j$ in time $t$ and with $q$ CPA queries to $Enc$ can break the CPA security of $\mathit{Enc}$ with $p=1/2+\varepsilon$. Thus the $1$-of-$n$ MFKDF construction w.r.t. $R$ is $(t,q,\varepsilon)$-secure w.r.t. $F_j$. \qed \newline

\noindent \textbf{\mbox{Theorem 4.}} Let $P$ be any policy $P \in \mathcal{P}(\mathcal{P}(S) \setminus \varnothing) \setminus \varnothing$. Let $E_i$ denote $\sum\{m_1,m_2,...,m_n\}$ for all ${C_i \in P}$. Let ${j=\mathit{argmin}_i E_i}$, and $F_P=\mathit{Stack}(C_j)$. Then the policy-based MFKDF construction of \S\ref{sec:policy_mfkdf_construction} w.r.t $\mathit{HKDF}$ and $P$ shall be \mbox{$(t,q,\varepsilon)$-secure} w.r.t. $F_P$. \newline

\noindent \textit{Proof.} For all $C_i \in P$, let $F_i = \mathit{Stack}(C_i)$. Per Cor. 2.1, $F_i$ is $m_i$ entropy secure where $m_i=E_i=\sum m_j$ for $m_j \in C_i$. Let $R = \{F_1, F_2, ..., F_n\}$ where $F_i$ is $m_i$-entropy secure. Now, the policy-based MFKDF w.r.t. $P$ is the $1$-of-$n$ MFKDF w.r.t. $R$. Thus per Thm. 3, it is \mbox{$(t,q,\varepsilon)$-secure} w.r.t. $F_j=F_P=\mathit{Stack}(C_j)$ where $j=\mathit{argmin}_i m_i$. \qed \newline

\noindent \textit{Comment.} We have now shown that given a policy $P$, the policy-based MFKDF is secure w.r.t. $\mathit{Stack}(C)$, where $C$ is the allowable combination in $P$ with lowest total entropy. Per Cor. 2.1, $\mathit{Stack}(C)$ is a computational ($\sum m_i$)-entropy factor where ${F_i \in C}$ are $m_i$-entropy factors. Therefore, we have achieved our entropy security goal of \S\ref{sec:setup}; the remaining goals follow directly from $\mathit{HKDF}$. \newline

\end{document}